\documentclass[manuscript,nonacm]{acmart}

\AtBeginDocument{%
  }

\usepackage{amsmath}
\usepackage[normalem]{ulem} %
\usepackage{enumitem}
\usepackage{hyperref}
\usepackage{graphicx}

\setcopyright{none}

\usepackage{array}
\usepackage{multirow}
\usepackage{ragged2e}
\usepackage{multicol}
\usepackage{color}

\usepackage{tikz}

\usepackage{colortbl}

\usepackage{multirow}
\usepackage{tabularx}
\usepackage[linesnumbered,ruled,vlined]{algorithm2e}
\usepackage{makecell}
\usepackage{booktabs}
\usepackage{soul}
\usepackage{changepage}
\usepackage{threeparttable}
\usepackage[table]{xcolor}

\usepackage{textcomp}
\usepackage{bbding}

\usepackage{comment}
\usepackage{makecell}
\usepackage{mfirstuc}
\usepackage{xurl}
\usepackage{pifont}
\usepackage{caption}
\usepackage{subcaption}
\usepackage[many]{tcolorbox}
\usepackage{hyperref}

\usepackage{tikz}
\usepackage{geometry}
\usepackage{wrapfig}

\newcommand{\etal}{{\em et al.}\xspace}
\newcommand{\eg}{{\em e.g.,}\xspace}
\newcommand{\ie}{{\em i.e.,}\xspace}

\usepackage{tikz}
\usepackage{adjustbox}
\usetikzlibrary{positioning, shapes, arrows.meta}

\newcommand{\OA}[1]{\textcolor{black}{#1}} %

\usepackage{rotating} %

\makeatletter
\@ACM@screentrue
\hypersetup{
    colorlinks=true,
    linkcolor=blue,
    citecolor=blue,
    urlcolor=blue
}
\makeatother

\begin{document}

\begin{center}
    \large\textcolor{red}{\textbf{Preprint. Under review at ACM Computing Surveys.}}
\end{center}

\title[\OA{The Evolution of Binary Decompilation in the Modern
Era}]{The Evolution of \OA{Binary} Decompilation in the Modern Era: A Taxonomy, Literature Review, and Future Perspectives}

\author{Omar Abusabha}
\affiliation{%
  \institution{Sungkyunkwan University}
  \city{Suwon}
  \country{South Korea}}
\email{404970@g.skku.edu}

\author{Sungjae Hwang}
\affiliation{%
  \institution{Sungkyunkwan University}
  \city{Suwon}
  \country{South Korea}}
\email{sungjaeh@skku.edu}

\begin{abstract}

Decompilation has become a foundational technique in software engineering and security analysis, and it is now advancing through the integration of modern machine learning (ML) approaches. This article presents a systematic review of 
\OA{66}
decompilation studies published over the past decades and develops a comprehensive taxonomy of methodologies employed in contemporary research.
We further examine trends in evaluation metrics, tools, and benchmarks used to assess state-of-the-art approaches. Our review reveals key challenges, such as the lack of reliable ground truth and the absence of standardized benchmarks, which hinder rigorous comparison. Finally, we outline future research directions.

\end{abstract}


\maketitle

\section{Introduction}

Binary decompilation (\OA{a.k.a. reverse} compilation)—the process of
translating low-level machine code back
into a high-level, source-like representation—plays a
pivotal role in modern program analysis. 
As source code is often unavailable due to
distribution practices or proprietary restrictions, 
decompilers offer a critical means for
recovering program semantics. 
These reconstructed representations underpin 
a wide range of downstream tasks,
including but not limited to malware analysis~\cite{bruschi2006using, saidi2010experiences, mirzaei2021scrutinizer, botacin2019revenge, vdurfina2013psybot}, 
binary translation~\cite{myreen2008machine, myreen2012decompilation}, 
porting~\cite{verbeek2020sound}, 
and code similarity analysis~\cite{wang2023decompilation, caliskan2015coding}.
They also aid in
vulnerability detection~\cite{mantovani2022convergence, han2023queryx, reiter2024automatically}, 
exploitation~\cite{han2023queryx}, patching~\cite{reiter2024automatically}, 
validation~\cite{kim2022reverse, tychalas2021icsfuzz, tsang2024ffxe, wu2024your}
\OA{and root cause 
analysis~\cite{park2024benzene}.
}

Decompilation research has roots in the 1960s, beginning with
Maurice Halstead's work
on the D-Neliac decompiler~\cite{halstead1962machine}.
\OA{This work} demonstrated the feasibility
of transforming machine code into human-readable,
source-like representations.
In the 1970s, \OA{Hollander~\etal~\cite{hollander1974syntax}
formalized decompilation into distinct stages, including}
initialization, scanning, parsing, construction, and
code generation.
Concurrently, other studies explored the usability
and practical deployment of
decompilers~\cite{housel1974methodology, weller1974pragmatic}.
By the 1990s, Cifuentes~\etal~\cite{cifuentes1994reverse}
introduced a compiler-inspired pipeline
organized into front-end, middle-end,
and back-end phases, establishing the
architectural basis of many modern
\OA{traditional} decompilers.
\OA{Accordingly, we define the
\textit{modern era} of binary decompilation
as beginning in the 1990s, when decompilation
evolved toward systematic, multi-stage
frameworks for recovering source-level
information from binaries.
This era differs from earlier research,
which primarily established the feasibility,
fundamental stages, and practical foundations
of decompilation, by progressively advancing
the architectures and techniques used to
recover source-level information.}
\OA{A subsequent shift occurred in 2018,
when} Katz~\etal~\cite{katz2018using}
reimagined decompilation
\OA{as a translation task, marking the
emergence of modern neural decompilers.
}

\begin{figure*}[t!]
    \centering
        \includegraphics[width=0.65\linewidth]{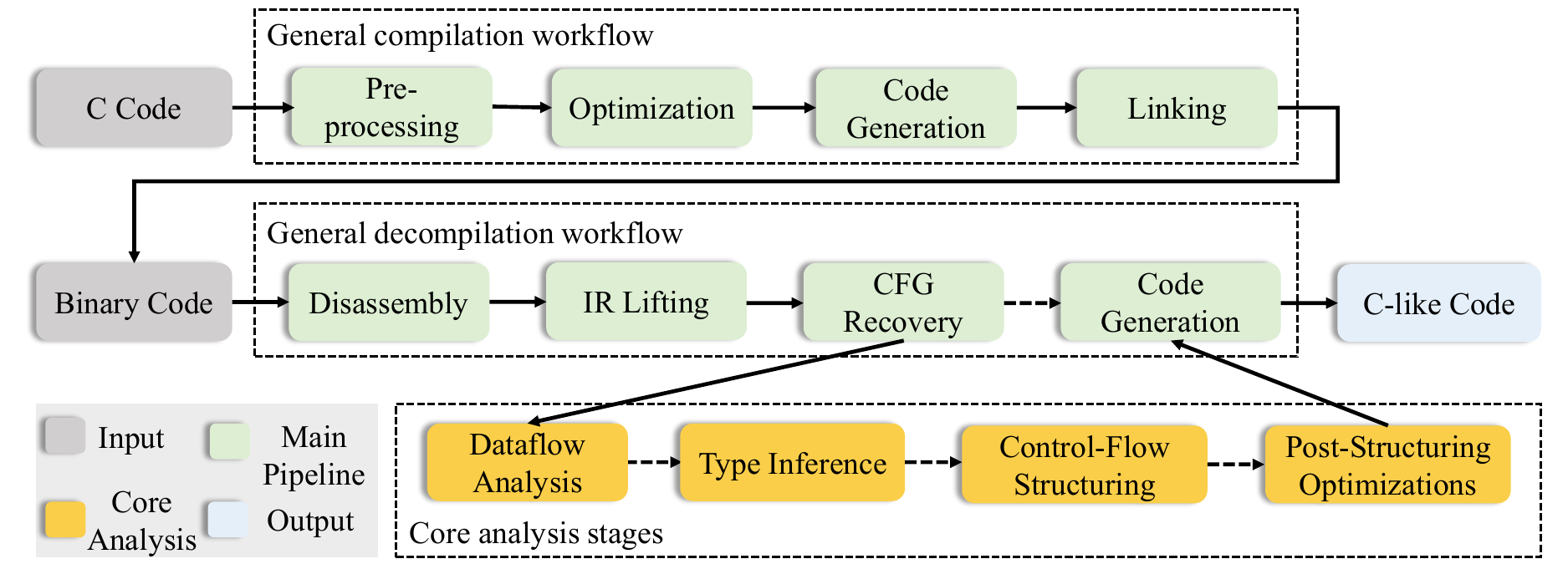}
        \caption{
            \OA{
                Overview of modern compilation and decompilation workflows.
            }
            \OA{
                We use directional 
                dependencies to indicate these relationships
                without implying a strictly
                sequential pipeline.
            }
       }
    \label{fig:full_framework}
\end{figure*}

Figure~\ref{fig:full_framework} 
provides \OA{an overview of the modern 
compilation and decompilation workflows.} 
\OA{Modern compilers (\eg GCC~\cite{gcc} and Clang~\cite{clang})
translate source code into executable binaries 
through multiple stages, including preprocessing 
(\eg semantic and syntactic validation), 
intermediate representation (IR) generation, 
optimization (\eg inlining), code generation, and linking, 
ultimately producing an executable binary.
}
\OA{To reverse this process, decompilers adopt 
different recovery workflows.}
Traditional decompilers (\eg Hex-Rays~\cite{hexrays_decompiler} 
and Ghidra~\cite{ghidra}) follow a staged architecture, 
starting with disassembly and IR lifting 
(\eg Microcode and P-code, respectively), 
\OA{
proceeding through control-flow graph (CFG) recovery, 
data-flow analysis, type inference, 
control-flow structuring, and post-structuring optimizations, 
and generating high-level source-like code.
}
In contrast, neural decompilers 
(\eg BTC~\cite{hosseini2022beyond}, 
Neutron~\cite{liang2021neutron}) 
\OA{use learned neural models as 
the primary mechanism 
for recovering source-level 
representations directly 
from binary instructions.
}
\OA{Moreover,} while compilers optimize for execution
performance and binary size, decompilers instead prioritize
\OA{semantic recovery, readability, and structural clarity.}

Despite decades of progress, decompilation 
remains a
challenging \OA{process} due to the lossy 
nature of compilation,
which discards semantic cues such as variable types,
function boundaries, and control structures.
These challenges are further amplified by compiler diversity,
platform-specific conventions, and
intentional obfuscation.

\OA{Modern binary decompilation has evolved beyond
a single reconstruction problem into a collection of
interdependent tasks, each supported by different
architectural and methodological choices.
However, these developments remain scattered across
studies of individual tasks, systems, and evaluation
methods, making it difficult to understand how
decompilation techniques have evolved, how their
design choices affect the recovered code, and which
directions have been explored to improve decompilation.
}
To fill this gap, we conduct a
Systematic Literature Review (SLR)~\cite{keele2007guidelines}
focused on C-language binaries.
\OA{Accordingly, we} formulate
five primary research questions.

\begin{itemize} 
\item \textbf{RQ1:} \OA{What architectural paradigms have shaped decompilation systems?} (\autoref{sec_full_dec})

\item \textbf{RQ2:} \OA{What task-oriented techniques have shaped decompilation?} (\autoref{sec_task_dec})

\item \textbf{RQ3:} \OA{How has the effectiveness of full-pipeline decompilers been evaluated?} (\autoref{sec_analysisn_trends})

\item \textbf{RQ4:} \OA{What capabilities and metrics are used to evaluate task-oriented decompilation techniques?} (\autoref{sec_mtrics_trends})

\item \textbf{RQ5:} \OA{What tools, frameworks, and benchmarks support decompilation research and evaluation?} (\autoref{sec_implementation_trends})
\end{itemize}

The remainder of this paper is organized as follows.
\OA{\autoref{sec_research_meth} describes the review methodology
adopted in this study.
}
\OA{\autoref{sec_full_dec} reviews the architectural paradigms that
have shaped full-pipeline decompilation systems, while
\autoref{sec_task_dec} discusses task-oriented decompilation
techniques.
}
\OA{\autoref{sec_analysisn_trends} examines how the effectiveness of
full-pipeline decompilers has been evaluated, while
\autoref{sec_mtrics_trends} reviews the capabilities and metrics
used to evaluate task-oriented decompilation techniques.
}
\OA{
\autoref{sec_implementation_trends} surveys the tools, frameworks,
and benchmarks that support decompilation research and evaluation.
}
\autoref{sec_challlenges} outlines open challenges
and future research directions, while
\autoref{sec_threat} addresses threats to validity.
Finally, \autoref{sec_conclusion} concludes the paper.

\section{Research Methodology}\label{sec_research_meth}

\begin{figure*}[htbp]
    \centering
        \includegraphics[width=0.99\linewidth]{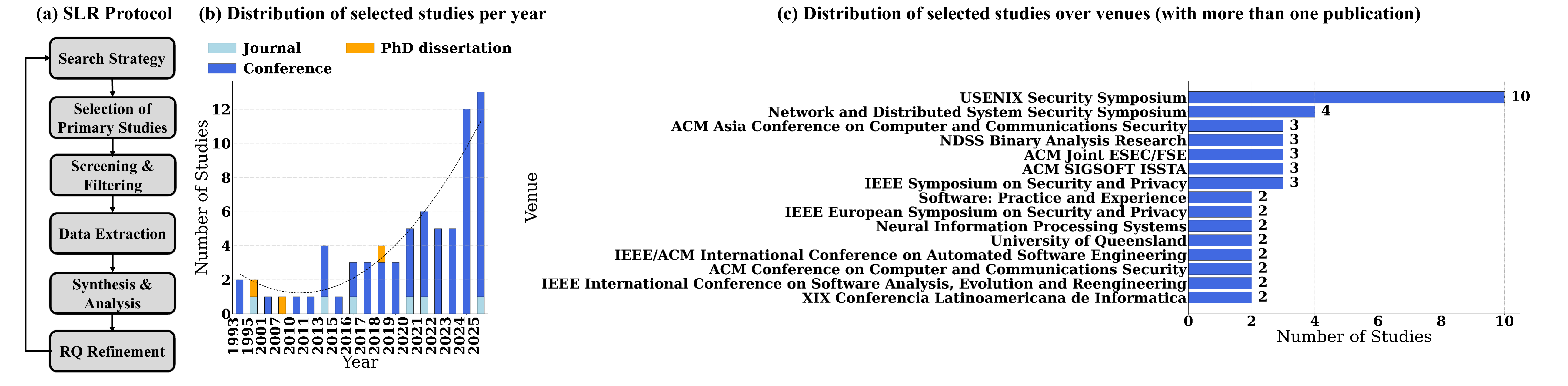}
        \caption{
            Overview of \OA{binary} decompilation research, 
            including \OA{(a) the 
            Systematic Literature Review (SLR) methodology
            adopted in this study following the 
            guidelines by Kitchenham \etal~\cite{keele2007guidelines}, 
            (b) the annual growth of selected studies since 1993, and 
            (c) peer-reviewed publication venues 
            with at least two selected studies.
            }
        }
    \label{fig:timeline_citation}
\end{figure*}

\OA{
In this section, we describe the methodology used to identify major 
research trends and trace the historical evolution of binary 
decompilation.
}
To ensure methodological rigor, 
we followed the SLR
guidelines by
Kitchenham \etal~\cite{keele2007guidelines}.
\OA{
The overall process is 
illustrated in~\autoref{fig:timeline_citation}(a), 
described next.
}

\subsection{Search Strategy}

\OA{ At this stage, we define the 
academic sources and search strings used to identify 
relevant studies on binary decompilation. 
We combine broad literature searches with 
targeted queries to improve coverage across the
different decompilation paradigms 
and tasks. 
}

\subsubsection{Search \OA{Sources}} 
\emph{Google Scholar} \OA{served as the} 
primary search engine due to
its broad indexing 
\OA{coverage across publishers and disciplines, including}
gray literature, 
which enhanced recall.
To ensure coverage of high-quality peer-reviewed sources, 
we supplemented the search with 
\emph{IEEE Xplore} and
the \emph{ACM Digital Library}.

\subsubsection{Search Strings}
We \OA{initially identified relevant domain terms
from foundational literature to construct
comprehensive search queries.}
\OA{For example, foundational studies on
reverse compilation~\cite{cifuentes1994reverse},
SSA-based decompilation~\cite{van2007static},
and human-centric binary-code
decompilation~\cite{yakdan2018human}
helped identify initial terms such as
\textit{``decompiler''},
\textit{``type inference''}, and
\textit{``control-flow structuring''}.} 
\OA{We then expanded and iteratively refined
the keyword set, reducing the initial set of
210 candidate keywords to a final set of
64 decompilation-specific keywords.}

\begin{table}[ht]
\centering
\caption{Inclusion and Exclusion Criteria.}
\resizebox{0.99\linewidth}{!}{

\begin{tabular}{p{18cm}}
\toprule
\textbf{Inclusion Criteria} \\
\midrule
\textbf{I1:} Papers published in peer-reviewed venues (\ie conferences or journals) \\
\textbf{I2:} Papers focusing on decompilation of binaries originally written in C \\
\textbf{I3:} \OA{Papers addressing either full-pipeline decompilation or task-oriented decompilation techniques covered by our taxonomy: type recovery, control-flow structuring, expression simplification, compiler idiom recovery, function recovery, variable name recovery, code summarization, and code refinement} \\
\midrule
\textbf{Exclusion Criteria} \\
\midrule

\textbf{E1:} Non-peer-reviewed publications, including arXiv preprints, keynotes, editorials, tutorials, panel discussions,
\OA{blog posts, practitioner articles, and engineering write-ups}
\\
\textbf{E2:} \OA{Survey, SoK, position, or taxonomy papers, unless used only for background discussion rather than inclusion in the primary study set (\eg disassembly~\cite{pang2021sok}, binary lifting~\cite{liu2022sok}, CFG recovery~\cite{wang2023survey}, or function identification~\cite{koo2021look})} \\

\bottomrule
\end{tabular}

}
\label{tab:selection_criteria}
\end{table}

\subsection{Selection of Primary Studies} %

\OA{
We selected the primary studies through
iterative screening of titles, abstracts, and full texts,
according to the inclusion and exclusion criteria defined in
\autoref{tab:selection_criteria}.
}

\subsubsection{\OA{Screening of Retrieved Studies}}
\OA{We manually screened over 120 studies
retrieved through the search queries to determine
their relevance to binary decompilation.
For instance, broad terms such as
\textit{``reverse engineering''} and
\textit{``binary analysis''} were retained
in the final keyword set to improve recall.
Although these terms retrieved a larger
number of studies, they also returned work
focused on general binary analysis
and other adjacent topics.}

\subsubsection{Inclusion-Exclusion Criteria}

To maintain the focus and relevance of our review, we applied the 
inclusion and exclusion criteria listed in~\autoref{tab:selection_criteria}.
We included peer-reviewed studies (I1) that focus on decompiling 
binaries originally written in C (I2).
Eligible studies address core decompilation challenges, such as 
type inference, control-flow structuring, and post-structuring 
optimization (I3).
Both traditional and learning-based approaches were considered.
\OA{To preserve methodological consistency, we excluded 
non-peer-reviewed sources, including arXiv preprints, tutorials, 
keynotes, blog posts, practitioner articles, and engineering 
write-ups.}
\OA{This also applies to practitioner-oriented materials for widely 
used decompilers, such as Ghidra~\cite{bulazel2019ghidra}, unless 
the relevant technique is documented in a peer-reviewed 
publication (E1).}
\OA{For this reason, we include Hex-Rays~\cite{guilfanov2008decompilers,guilfanov2018decompiler18} 
in our discussion of 
traditional decompiler systems where
its type-inference mechanism 
is described in 
peer-reviewed work~\cite{guilfanov2001simple}.
}
\OA{
Accordingly, for actively maintained tools,
our analysis reflects the designs 
and capabilities documented
in the surveyed literature 
at the time of publication and does
not necessarily represent their
current implementations.
}
\OA{In addition,}
\OA{
We also excluded survey, SoK (Systematization of Knowledge),
position, and taxonomy papers from the primary study set,
unless they were used solely for background discussion
such as disassembly~\cite{pang2021sok}, binary 
lifting~\cite{liu2022sok}, CFG recovery~\cite{wang2023survey}, 
or function identification~\cite{koo2021look} (E2).}

\subsubsection{Distribution of Studies}

\autoref{fig:timeline_citation}(b) depicts 
the annual distribution of the
\OA{72}
primary studies included in 
our review
~\footnote{A full list of the 
papers is available
here: \url{https://github.com/lion10/csur2026_binary_decompilation.git}.
}.
\OA{
Among these studies, three are PhD dissertations, one is a SoK
paper, one is a survey, and one is an empirical benchmarking
study, which were used to support the review but were not
included as primary technique contributions.
}
Although decompilation efforts date back 
to the 1960s, the foundations of modern 
decompiler design emerged in the 1990s, 
which marks the starting point of our 
review focus.
Research activity remained limited until 
around 2013, followed by a steady upward 
trend and a sharp increase after 2020.
The highest volume of publications occurred 
in \OA{2025}, with \OA{13} peer-reviewed 
studies published that year, reflecting 
heightened research activity.
Most studies appeared in conference venues, 
which reflects the field's emphasis on 
timely dissemination, particularly in areas 
related to security, software analysis, and 
program understanding.

\subsubsection{\OA{Distribution of Publication Venues}}
\autoref{fig:timeline_citation}(c) shows the 
most common publication venues.
The majority of selected studies appeared in 
security-oriented venues, with USENIX Security 
(\OA{10} studies) being the most frequent venue, 
followed by \OA{the Network and Distributed System 
Security (NDSS) Symposium}.
Software engineering venues such as ESEC/FSE 
and ISSTA also appear, demonstrating the 
cross-disciplinary nature of decompilation 
research.

\subsection{\OA{Data Extraction \& Synthesis}}

\OA{
After selecting the 66 primary studies, 
we synthesized the
literature according 
to the five research questions.
}
\OA{For each study, we extracted its objective, decompilation 
task, technical approach, evaluated capability, metrics, 
benchmarks, and reported limitations.}
\OA{We then grouped related studies into consistent categories to 
identify historical trends, recurring design choices, evaluation 
patterns, and open challenges.}
As a final note, we provide a high-level 
\OA{organization of the field} and 
we recommend readers to refer to 
the original publications 
for an in-depth details.

\section{RQ1: Research Trends 
\OA{in Full-Pipeline Binary Decompilation}
}\label{sec_full_dec}

\begin{figure*}[t!]
    \centering
        \includegraphics[width=0.5\linewidth]{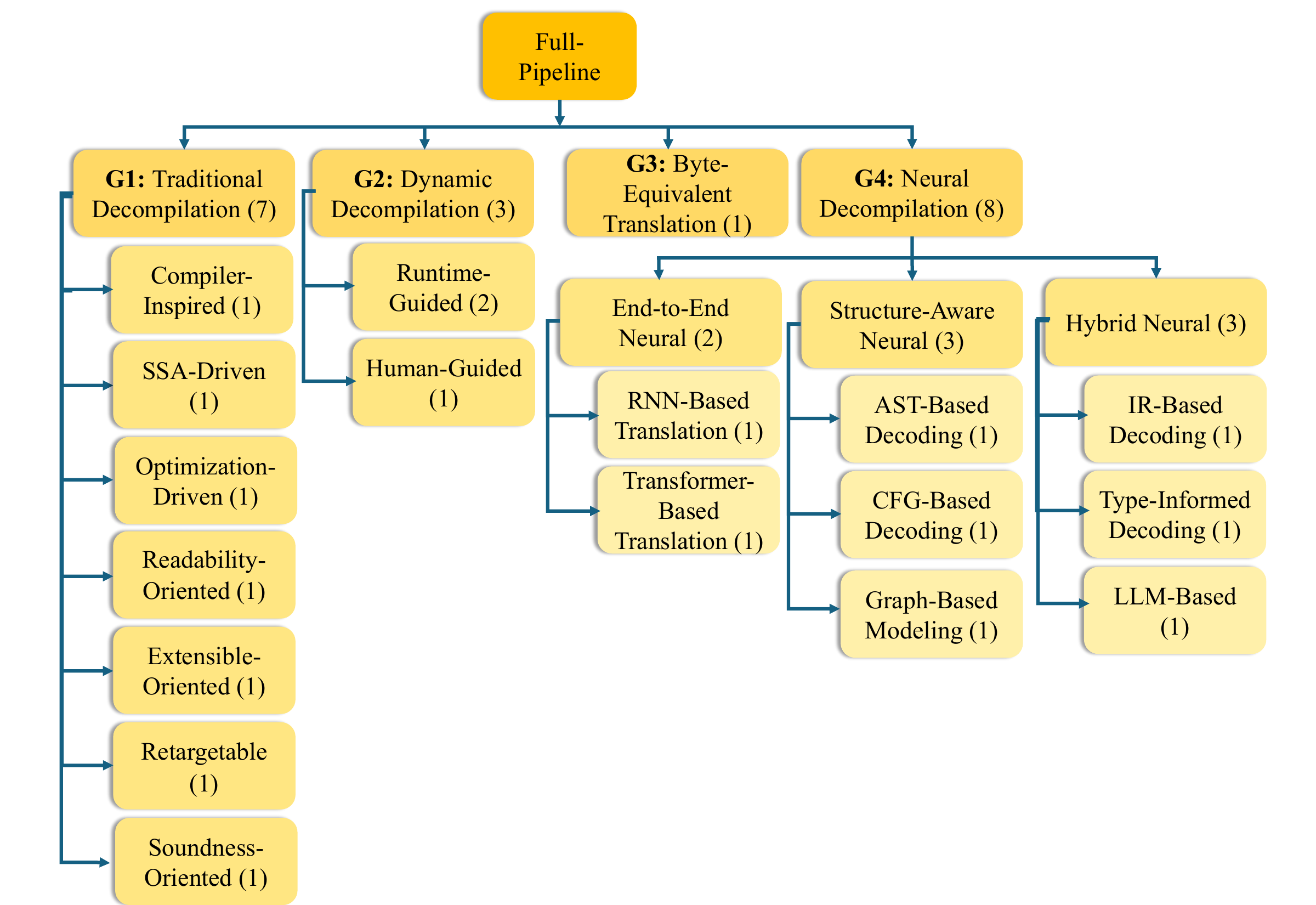}
        \caption{\OA{
        Taxonomy of full-pipeline 
        decompilation paradigms 
        across four generations.
        }}
    \label{fig:tax_full_pipline_arch}
\end{figure*}

\OA{
In this section, we review full-pipeline decompilation
approaches and organize them by their dominant architectural
paradigms.
\autoref{fig:tax_full_pipline_arch} highlights how decompiler
design has evolved across four generations (G\#), 
discussed next.
}

\subsection{\OA{G1: Traditional Decompilation}}
\OA{
Traditional decompiler architectures have evolved around
different design priorities over time.
We organize representative systems according to their dominant
architectural priority, while noting that individual systems may
combine characteristics from multiple designs, as described next.
}

\subsubsection{Compiler-Inspired}

\OA{
dcc~\cite{dcc} is an early modular decompiler
that applies a compiler-inspired architecture to
binary-to-C translation
~\cite{cifuentes1995decompilation,cifuentes1994reverse}.
It uses a dual-level IR comprising low-level Icode and
high-level Hcode, together with idiom recognition,
data-flow analysis, interval-based control-flow
structuring~\cite{cifuentes1993methodology}, and
library-signature matching.
This organization established a staged decomposition in which
low-level machine information is progressively transformed
into higher-level program representations.
}

\subsubsection{IR-Driven}

\OA{
Boomerang~\cite{boomerang} is a retargetable,
open-source decompiler that places Static Single
Assignment (SSA) form at the center of its
architecture~\cite{van2007static}.
Its analyses include expression propagation,
register-preservation analysis, sparse data-flow-based
type analysis, and modular front-end and back-end
components for supporting multiple targets
~\cite{emmerik2004using}.
Unlike the staged representations of dcc,
Boomerang relies on a common SSA-based representation
to support multiple downstream analyses.
}

\subsubsection{Optimization-Driven}

\OA{
Hex-Rays~\cite{hexrays_decompiler} is a commercial
decompiler that translates architecture-specific
instructions into a microcode IR and performs extensive
IR simplification before code generation
~\cite{guilfanov2001simple,
guilfanov2008decompilers,guilfanov2018decompiler18}.
Its analyses include constant and register propagation,
dead-code elimination, operand folding, high-level
control-flow structuring, and FLIRT-based library
recognition~\cite{hex_rays_flirt}.
Its architecture therefore emphasizes progressive
IR optimization before reconstructing high-level code.
}

\subsubsection{Readability-Oriented}

\OA{
C-Decompiler targets readable C output from compiled
programs~\cite{c_decompiler_conf,chen2013refined}.
It combines a custom IR with refined stack tracing,
parameter and local-variable recovery,
inter-basic-block register propagation,
function-signature recognition, and type and array
recovery.
These analyses primarily reduce redundant low-level
artifacts and reconstruct higher-level source
abstractions, making readability the dominant
architectural objective.
}

\subsubsection{Retargetable}

\OA{
RetDec~\cite{retdec}, originating from the Lissom
project~\cite{vdurfina2011design}, separates
target-specific instruction decoding from reusable
middle-end analyses.
It uses ISAC architecture descriptions to generate
instruction decoders, LLVM IR as a common intermediate
representation, idiom recognition
~\cite{kvroustek2013reconstruction}, control-flow
structuring, and IR optimization.
This separation allows the same decompilation pipeline
to be adapted across architectures with limited changes
to target-specific components.
}

\subsubsection{Extensibility-Oriented}

\OA{
REcompile~\cite{yakdan2013recompile} is built on
IDA Pro~\cite{ida_pro} as an extensible framework for
static binary analysis.
It combines an SSA-form IR with dead-code elimination,
expression simplification, parameter detection,
graph-based control-flow structuring, and basic-block
transformations for reducing \texttt{goto} statements.
Its modular design emphasizes the integration of new
analysis and transformation passes as binary-analysis
requirements evolve.
}

\subsubsection{Soundness-Oriented}

\OA{
FoxDec~\cite{foxdec} emphasizes sound,
semantics-preserving decompilation
~\cite{verbeek2020sound}.
Its architecture combines CFG recovery verified using
Isabelle/HOL, symbolic execution for aggregating
assembly instructions into higher-level constructs,
variable mapping between symbolic memory and C
variables, and type punning when precise source-level
types cannot be recovered.
Unlike architectures primarily optimized for
readability, FoxDec prioritizes verified transformations
and recompilable output.
}

\subsection{G2: Dynamic Decompilation}

\OA{
Dynamic decompilation refers to approaches in which
runtime evidence or analyst interaction guides the
decompilation process.
We group these approaches into runtime-guided and
human-guided decompilation, as discussed next.
}

\subsubsection{Runtime-Guided}
One of the earliest systems in this direction
was proposed by
Cifuentes \etal~\cite{cifuentes2001computer}, who integrated 
dynamic decompilation into the
UQBT framework~\cite{uqbt}. 
By executing binaries and observing runtime behavior, 
the system could address 
static challenges such as indirect jumps and
code-data interleaving. 
Building on this concept, Botacin \etal~\cite{botacin2019revenge} 
introduced the \textit{Revenge} framework,
combining interactive debugging (\eg via GNU Debugger) 
with decompilation. 
This allowed analysts to
iteratively examine and reconstruct
malware binaries,
yielding recompilable code that more accurately
reflects runtime semantics.
In these cases, the 
decompiled output encodes both
static structure and
analyst-driven interpretation.

\subsubsection{Human-Guided}
Burk \etal~\cite{burk2022decomperson} 
proposed \textit{Decompetition},
a system for iterative, human-guided decompilation.
The framework defines "perfect decompilation" as
the point at which recovered source code,
when compiled, yields a
byte-equivalent binary. 
Through incremental refinement and
real-time user feedback, 
the system allows reverse engineers 
to converge on minimal and semantically
accurate source representations. 
This process introduces an explicit stopping
criterion tied to correctness, 
rather than subjective readability alone.

\subsection{\OA{G3: Byte-Equivalent Translation Decompilation}}
This line of work reframes decompilation as a
search problem guided by
binary-level equivalence. 
Schulte \etal~\cite{schulte2018evolving} 
introduced BED, which applies
evolutionary algorithms to 
synthesize candidate programs from
a corpus of human-authored source code. 
Each candidate is compiled and evaluated 
using a byte-similarity fitness function, enabling
iterative refinement toward 
binary-equivalent outputs. 
While this method can recover
functionally accurate and structurally 
simplified code, its effectiveness is
limited by corpus coverage and
scalability across diverse input binaries.

\subsection{G4: Neural Decompilation}

Neural decompilation recasts binary-to-source
translation as a data-driven learning task, 
aiming to replace manual heuristics with
model generalization. While this shift
improves adaptability across architectures, it 
often reduces interpretability and remains 
limited by the quality and
coverage of training data.
\OA{
We organize neural decompilation into three primary
approaches, as described next.
}

\subsubsection{End-to-End Neural Models}

\OA{
Initial efforts applied neural machine translation (NMT)
to translate binary instructions directly into high-level
source code using learned representations rather than
architecture-specific heuristics.
}

\begin{itemize}

\item \textit{RNN-Based Translation.}  
Katz \etal~\cite{katz2018using} applied 
Recurrent Neural Networks (RNNs) 
in a sequence-to-sequence setup to 
translate short assembly snippets into C. 
\OA{
This demonstrated the feasibility of end-to-end neural
decompilation, but limited structural awareness and context
handling led to syntactic and semantic errors on longer
sequences.
}

\item \textit{Transformer-Based Translation.}  
Beyond The C (BTC)~\cite{hosseini2022beyond} 
advanced end-to-end learning by employing 
Transformer architectures
and Byte Pair Encoding (BPE) 
to support cross-language decompilation 
(\eg C, Fortran, Go, OCaml). 
\OA{
It treats binary and source code as plain text and improves
retargetability and language independence, but may produce
semantically inaccurate sketches requiring human correction.
}

\end{itemize}

\subsubsection{Structure-Aware Neural Models}

\OA{
To address the limited structural context of sequence-based
models, later approaches incorporated syntactic and semantic
structure into the decoding process.
}

\begin{itemize}

\item \textit{AST-Based Decoding.}  
Coda~\cite{fu2019coda}, though an end-to-end framework, 
introduced a two-phase decoding framework
involving code sketch generation 
followed by iterative error correction. 
\OA{
It combines a type-aware encoder with an AST-based decoder
and compiler-guided correction to preserve syntactic structure
and improve semantic consistency.
However, it remains limited by complex ISAs, long programs,
and static-library dependencies.
}

\item \textit{CFG-Based Decoding.}  
Neutron~\cite{liang2021neutron} advanced 
the readability and generalization of
decompiled output through a 
novel use of basic operations as atomic translation units.
\OA{
Its attention-based NMT model uses finer segmentation and
rule-assisted reconstruction to improve alignment between
low-level and high-level semantics.
However, it performs less effectively on optimized code and
user-defined types.
}

\item \textit{Graph-Based Modeling.}  
N-Bref~\cite{fu2020n} introduced a 
structural transformer architecture 
tailored for non-linear code representations. 
\OA{
It combines assembly, AST, and tree-structured representations
to model control- and data-flow dependencies and separates
type recovery from source-code generation.
This improves structural modeling but increases dependence on
accurate preprocessing such as AST and CFG recovery.
}

\end{itemize}

\subsubsection{Hybrid Neural Models}

\OA{
Hybrid models combine neural learning with symbolic or
intermediate representations to address compiler transformations
that are difficult for purely neural models.
}

\begin{itemize}

\item \textit{IR-Based Decoding.}  
NeurDP~\cite{cao2022boosting} introduced 
a hybrid approach that incorporated IRs 
into the learning process. 
\OA{
It uses a GNN to map low-level code to a high-level IR and
segments basic blocks using its Optimal Translation Unit
strategy.
This improves robustness to instruction reordering and
dead-code elimination, but is limited to relatively simple
operations.
}

\item \textit{Type-Informed Decoding.}
SLaDe~\cite{armengol2024slade} targets 
semantically correct decompilation 
of optimized binaries across multiple
Instruction Set Architectures (ISAs), 
including x86 and ARM. 
\OA{
It combines a Transformer with the PsycheC symbolic type
inference engine~\cite{melo2017inference,melo2020type}
and uses equivalence testing to improve functional correctness.
However, performance decreases on longer code and some
semantically similar constructs remain difficult to distinguish.
}

\item \textit{LLM-Guided Refinement.}
LLM4Decompile~\cite{tan2024llm4decompile} 
leverages LLMs alongside traditional 
decompilers to enhance both readability
and executability of decompiled code. 
\OA{
It supports both direct binary-to-source translation through
LLM4Decompile-End and refinement of Ghidra pseudo-code through
LLM4Decompile-Ref.
Its staged training improves compilability and executability,
but dependence on reliable disassembly and decompiler output
makes it vulnerable to binary obfuscation.
}

\end{itemize}

\begin{figure*}[t!]
    \centering
        \includegraphics[width=0.9\linewidth]{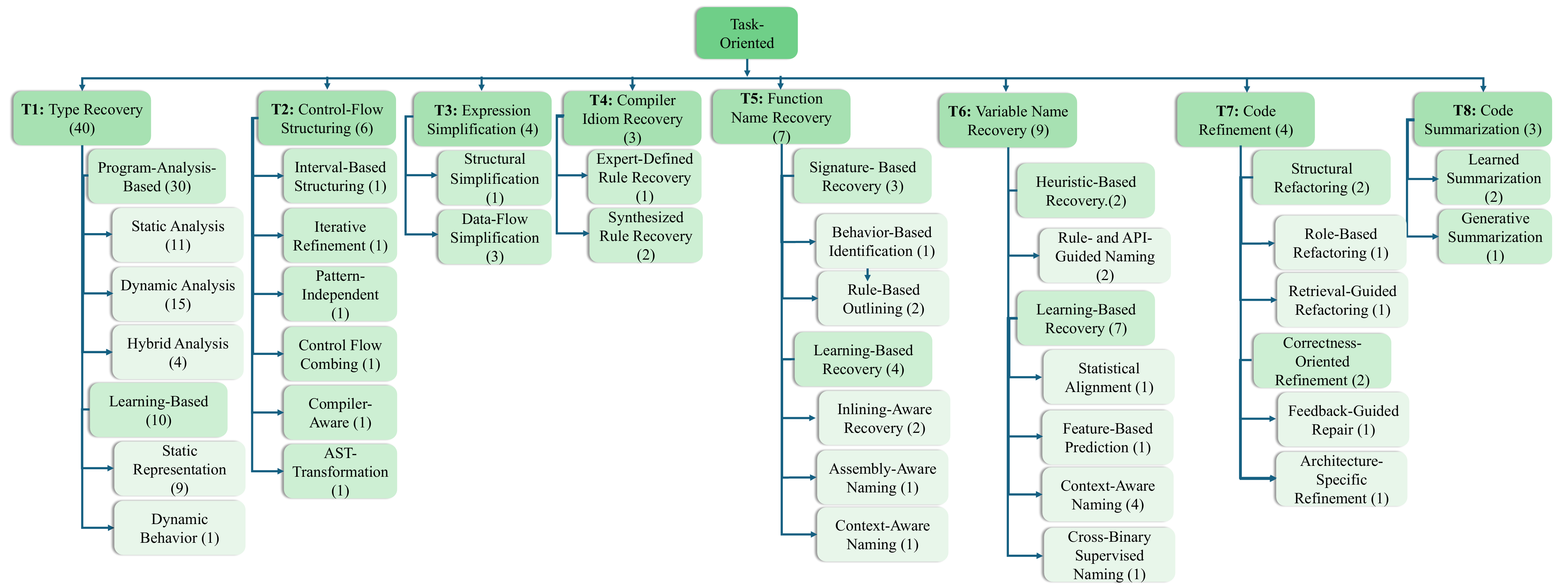} %
        \caption{
        \OA{
            Taxonomy of task-oriented 
            decompilation techniques 
            across eight tasks.
        }}
    \label{fig:task_oriented}
\end{figure*}

\section{\OA{RQ2: Research Trends in Task-Oriented Binary Decompilation}}
\label{sec_task_dec}

\OA{
Distinct from RQ1, we organize task-oriented decompilation research
into eight tasks, each targeting a specific aspect of source-code
recovery.
\autoref{fig:task_oriented} presents this taxonomy, with each task
denoted as T\#.
Generally speaking, 
Tasks 1--2 correspond to core decompiler stages,
Tasks 3--6 cover post-structuring optimizations,
and 
Tasks 7--8 operate mainly on decompiled code after code generation.
}

\subsection{T1: Type Recovery}

\OA{
Type recovery encompasses type inference, prediction, and
reconstruction of source-level type information that is
unavailable, incomplete, or unreliable.
}
Building on Caballero~\etal~\cite{caballero2016type}, 
who organized
38 type-inference \OA{systems} for C/C++ binaries, 
27 of which
target C specifically, 
we extend their taxonomy 
with \OA{13} additional
systems.
\OA{
We group these approaches 
into two categories,
described below.
}

\subsubsection{\OA{Program-Analysis-Based Approaches}}
\OA{
Early type-recovery research relied primarily on explicit program
analysis, deriving type information from constraints, data flow,
memory behavior, and runtime evidence.
These methods provide interpretable recovery rules and remain the
foundation of many decompilation pipelines.
}

\begin{itemize}

\item \textit{Static Analysis.}
\OA{Static approaches are primarily flow-based, deriving type
information from instructions, IRs, data flow, and program
constraints without executing the program.
Mycroft~\cite{mycroft1999type}, SECONDWRITE
~\cite{elwazeer2013scalable}, RHK~\cite{robbins2013theory},
and YM~\cite{yan2014conservative} recover scalar, pointer,
and function-related type relations through constraint solving.
Retypd~\cite{noonan2016polymorphic} and
PsycheC~\cite{melo2017inference,melo2020type} extend this process
with unification, subtyping, and polymorphic relations.
FFE~\cite{lim2006extracting}, DIVINE
~\cite{balakrishnan2007divine}, and
TDA~\cite{troshina2010reconstruction} reconstruct composite
types and data structures from memory-access and control-flow
relations.
OSPREY~\cite{zhang2021osprey} handles uncertain type evidence
probabilistically, whereas Trex~\cite{bosamiya2025trex} uses
formal reasoning to reconstruct types consistent with observed
machine behavior.
Static flow-based analysis provides broad program coverage and
interpretable type relations, but remains sensitive to compiler
optimization, aliasing, and incomplete interprocedural analysis.}

\item \textit{Dynamic Analysis.}
\OA{Dynamic approaches use either flow-based or value-based
analysis of concrete runtime behavior.
Flow-based approaches track type information through execution
traces, taint propagation, and memory accesses.
REWARDS~\cite{lin2010automatic},
POINTERSCOPE~\cite{zhang2012identifying},
UNDANGLE~\cite{schwarz2018automated}, and
TOP~\cite{zeng2013obfuscation} infer pointer and value types by
tracking runtime propagation.
DYNCOMPB~\cite{guo2006dynamic} unifies types associated with the
same runtime entities, while ARTISTE
~\cite{caballero2012artiste} and HOWARD
~\cite{slowinska2011howard} reconstruct dynamic objects,
recursive structures, and nested types.
Value-based approaches instead inspect concrete memory values,
layouts, and recurring patterns.
POLYGLOT~\cite{caballero2007polyglot},
AUTOFORMAT~\cite{lin2008automatic},
TUPNI~\cite{cui2008tupni}, DISPATCHER
~\cite{carbone2009mapping}, DC~\cite{dolgova2008automatic},
and DDE~\cite{slowinska2010dde} recover data formats and aggregate
layouts from observed memory behavior.
MEMPICK~\cite{haller2013mempick} classifies memory regions from
concrete values, whereas RDS~\cite{raman2005recursive} recovers
recursive data structures through memory-shape analysis.
Dynamic flow-based analysis provides precise evidence about
observed value propagation, whereas value-based analysis is
particularly effective for aggregate and recursive structures.
Both, however, remain limited by executed paths, selected inputs,
and visible runtime states.}

\item \textit{Hybrid Analysis.}
\OA{Hybrid approaches combine static program relations with
runtime evidence through flow-based, value-based, or combined
flow- and value-based analysis.
In flow-based analysis, TIE~\cite{lee2011tie} uses value-set
analysis and lattice-based solving to recover scalar and
pointer-related type relations, while
WCKK~\cite{wondracek2008automatic} uses memory-access patterns
to reconstruct nested data structures.
In value-based analysis, DDT~\cite{jung2009ddt} recovers memory
layouts from recurring value patterns across executions.
BCR~\cite{caballero2010binary} combines flow-sensitive
constraints, memory values, and repeated executions, representing
a flow- and value-based approach that recovers both type relations
and aggregate structures.
Hybrid analysis reduces ambiguity by combining complementary
evidence sources, but increases analysis complexity and remains
dependent on representative executions.}

\end{itemize}

\subsubsection{Learning-Based Approaches}
\OA{
More recent work shifts from manually specified inference rules
toward learned representations that capture recurring type
patterns from large collections of binaries and decompiled code.
These approaches complement program analysis by modeling semantic
regularities that are difficult to encode explicitly.
}

\begin{itemize}

\item \textit{Static Representations.}
\OA{Static learning-based approaches predict types from engineered
features, token sequences, or program graphs without requiring
runtime execution.
Debin~\cite{he2018debin} and
TypeMiner~\cite{maier2019typeminer} use engineered features to
predict variable and symbol types.
EKLAVYA~\cite{chua2017neural} predicts function signatures from
instruction sequences, while
DIRTY~\cite{chen2022augmenting},
Stir~\cite{peng2023statistical}, and
ReSym~\cite{xie2024resym} predict variable types and related
source-level information from ill-formed or decompiled-code
sequences.
TYGR~\cite{zhu2024tygr}, DRAGON~\cite{stewartdragon}, and
GENNM~\cite{xu2025unleashing} use graph representations to predict
types from structural and data-flow relations.
These approaches can recover patterns not captured by explicit
constraints, but their accuracy depends on training data,
representation quality, and generalization to unseen binaries.}

\item \textit{Dynamic Behavior.}
\OA{Dynamic learning-based approaches predict types from execution
traces, runtime operands, and observed program states.
StateFormer~\cite{peng2023statistical} models operand behavior
across executions to infer runtime type information.
These approaches learn semantic patterns from concrete behavior,
but require representative traces and incur runtime
instrumentation overhead.}

\end{itemize}

\subsection{T2: Control-Flow Structuring}

\OA{
Control-flow structuring transforms a recovered CFG into
high-level constructs through semantics-preserving
transformations.
}
\OA{
Behner~\etal~\cite{behner2025sok}
recently systematized control-flow 
structuring more broadly
in binary decompilation, while we exclude
compiler-side methods
~\cite{erosa1994taming,engel2011enhanced}.
We group existing approaches into six categories, 
described below.
}

\subsubsection{Interval-Based Structuring}
\OA{Cifuentes \etal~\cite{cifuentes1993structuring}
introduced an interval-based approach in
dcc~\cite{dcc}.
The method recursively partitions the CFG into maximal
single-entry regions and collapses recognizable loops and
conditionals into structured constructs.
Dominance and path analysis guide this process, while heuristics
handle irregular flows such as abnormal exits.
This provides a systematic foundation for CFG structuring but is
less effective for irreducible control flow.
}

\subsubsection{Iterative Refinement}
\OA{
To better handle irreducible regions,
Schwartz \etal~\cite{brumley2013native}
introduced Phoenix on top of BAP~\cite{bap}.
Phoenix separates cyclic and acyclic regions and temporarily
removes problematic edges through virtualization so that
structuring can proceed iteratively.
Removed edges are later represented using constructs such as
\texttt{goto}, \texttt{break}, or \texttt{continue} to preserve
semantics.
This improves coverage of irregular CFGs but may still retain
unstructured control when complete structuring is not possible.
}

\subsubsection{Pattern-Independent Structuring}
\OA{
Yakdan \etal~\cite{yakdan2015no}
introduced DREAM, built on top of
IDA Pro~\cite{ida_pro}. to avoid dependence on fixed structuring
templates.
Instead, DREAM uses topological ordering, dominance analysis,
conditional refinement, and rule-based inference to restructure
cyclic and acyclic regions without relying on predefined control
patterns.
This enables broader elimination of unstructured jumps, although
the resulting structure depends on the quality of the recovered
CFG and applied refinement rules.
}

\subsubsection{Control-Flow Combing}
\OA{
Gussoni \etal~\cite{gussoni2020comb}
introduced Combing, built on top of
rev.ng~\cite{revng}, to reduce the excessive nesting and duplication
that may result from aggressive \texttt{goto} elimination.
The method first decomposes the CFG into nested DAGs, restructures
overlapping control paths into diamond-shaped regions, and then
matches the resulting graph to C constructs such as
\texttt{if}, \texttt{switch}, and loop statements.
This produces structured control flow while controlling code
complexity, but requires substantial CFG transformation before
high-level reconstruction.
}

\subsubsection{Compiler-Aware Deoptimization}
\OA{
Basque \etal~\cite{basque2024ahoy}
introduced SAILR to distinguish control-flow artifacts introduced
by compiler optimizations from structures that may reflect
developer intent.
Built on angr~\cite{angr}, SAILR identifies patterns produced by
compiler transformations such as jump threading and switch
lowering and applies corresponding deoptimizations to recover
higher-level control flow.
This compiler-aware strategy avoids indiscriminate
\texttt{goto} elimination, but its effectiveness depends on the
compiler transformations that can be recognized.
}

\subsubsection{\OA{AST-Level Transformation}}
\OA{
Enders \etal~\cite{enders2025jump}
proposed an AST-level approach for recovering switch statements
on top of Ghidra~\cite{ghidra}.
Rather than relying on explicit jump tables or fixed CFG patterns,
the method identifies clusters of comparisons over the same
selector expression and rewrites equivalent
\texttt{if}-\texttt{else} structures into
\texttt{switch} statements.
This allows recovery across different switch-lowering patterns,
but the evaluation is limited to GCC binaries and
\texttt{-O0}.
}

\subsection{T3: Expression Simplification}

\OA{
Expression simplification rewrites 
low-level expressions,
redundant temporaries, and 
opaque predicates
into simpler forms to reduce 
syntactic complexity while
preserving program semantics.
We group existing approaches into two categories, 
described below.
}

\subsubsection{\OA{Structural Simplification}}

\OA{
Early work focused on removing low-level structural artifacts
introduced during compilation or obfuscation.
Saidi \etal~\cite{saidi2010experiences} proposed transformations
including stack normalization, function-body reassembly, and
calling-convention rewriting.
These transformations remove artifacts such as junk instructions
and semantic no-ops and reconstruct more coherent local code
regions.
Their main advantage is the direct removal of machine-level
artifacts, although they provide limited support for simplifying
more complex expressions and predicates.
}

\subsubsection{\OA{Data-Flow Simplification}}

\OA{
Data-flow approaches use intermediate representations, often
SSA-based, to propagate expressions and eliminate redundant
computations.
REcompile~\cite{yakdan2013recompile}, for example, performs
expression propagation and dead-code elimination while preserving
global variables to reduce the risk of semantic loss.
DREAM++~\cite{yakdan2016helping} further applies constant
propagation and common subexpression elimination, together with
symbolic reasoning through Z3~\cite{Z3}, to simplify complex
expressions and predicates.
dewolf~\cite{enders2022dewolf} extends this direction with
congruence analysis, a custom graph-based logic engine, and
pointer transformation.
Congruence analysis merges semantically equivalent variables or
expressions, while pointer transformation rewrites low-level
pointer arithmetic into higher-level constructs such as array
indexing and reference-based access.
These techniques can simplify expressions beyond conventional
propagation and elimination, but richer symbolic and graph-based
reasoning may increase computational cost and remains dependent
on accurate IR and data-flow information.
}

\subsection{\OA{T4: Compiler Idiom Recovery}}

\OA{
Compiler idiom recovery identifies optimized instruction
sequences that encode high-level operations and reconstructs
their original semantics.
We group existing approaches into two categories, 
described below.
}

\subsubsection{\OA{Expert-Defined Pattern Recovery}}

\OA{
Expert-defined approaches use manually crafted patterns and
static analysis to recognize known compiler idioms.
Lissom~\cite{kvroustek2013reconstruction}, for example,
matches predefined instruction sequences in LLVM IR to recover
high-level expressions.
Operating on an IR can abstract architecture-specific instructions
and allow rules to be reused across architectures.
However, recognition remains dependent on manually maintained
patterns, and verbose IR sequences can make matching difficult,
particularly when compiler optimizations or instruction semantics
vary substantially.
}

\subsubsection{\OA{Synthesized Pattern Recovery}}

\OA{
Synthesized approaches automate idiom discovery to reduce the
manual effort required to construct pattern databases.
PIdARCI~\cite{enders2021pidarci} generates instruction sequences
from program templates across different constant values, compiler
versions, and optimization settings, then anonymizes and clusters
them to identify recurring patterns.
This improves coverage and facilitates updates for evolving
compiler toolchains.
However, transformation rules for recovering original constants
still require manual derivation, and abstraction during
anonymization may lose distinctions between semantically different
instruction forms.
dewolf~\cite{enders2022dewolf} integrates PIdARCI within its
decompilation pipeline for idiom recovery and expression
simplification.
}

\subsection{\OA{T5: Function Name Recovery}}
\OA{
Function-name recovery reconstructs function identities or
descriptive identifiers that are unavailable in stripped or
optimized binaries.
We group existing approaches into two categories, 
described below.
}

\subsubsection{Signature- and Analysis-Based Recovery}

\OA{
These approaches recover known function identities or abstractions
using predefined signatures, behavioral evidence, and
program-analysis rules.
}

\begin{itemize}

\item \textit{\OA{Behavior-Based Identification.}}
\OA{
UNSTRIP~\cite{jacobson2011labeling} constructs semantic
descriptors from system calls and their arguments using CFG
analysis and backward slicing.
Because these behaviors can remain stable despite code-layout
changes, the approach is more robust than exact signature matching
to some compiler transformations, but is mainly applicable to
wrapper functions interacting with the operating system.
}

\item \textit{\OA{Rule-Based Outlining.}}
\OA{
DREAM~\cite{yakdan2015no} and
DREAM++~\cite{yakdan2016helping}
identify known library-function bodies embedded within larger code
regions and replace them with corresponding function calls.
This restores known modular abstractions but remains limited to
function bodies that can be reliably recognized.
}

\end{itemize}

\subsubsection{Learning-Based Recovery}

\OA{
These approaches infer function-level semantics from learned code
representations rather than relying solely on predefined
signatures or analysis rules.
}

\begin{itemize}

\item \textit{\OA{Inlining-Aware Recovery.}}
\OA{
FUNCRE~\cite{ahmed2021learning} uses a RoBERTa-based model trained
on disassembled traces aligned with source-level markers to
identify inlined library functions in optimized binaries.
It supports recovery when inlining obscures conventional
signatures, but its accuracy decreases when multiple functions are
inlined into the same region.
ERASE~\cite{zhang2024optimizing} further addresses inlining-induced
modularity loss using LLM-guided analysis, recursive symbolic
execution, and functional similarity to identify redundant inlined
function bodies and restore function calls.
This can recover higher-level modular structure, but depends on
accurate control- and data-flow information from earlier
decompilation stages.
}

\item \textit{\OA{Assembly-Aware Naming.}}
\OA{
AsmDepictor~\cite{kim2023transformer}
formulates function naming as translation from assembly
instructions to descriptive identifiers.
It uses a Transformer with byte-pair encoding to capture
instruction-level semantics and reduce vocabulary sparsity.
This enables naming beyond known signatures but depends on
representative training data and generalization to unseen code.
}

\item \textit{\OA{Context-Aware Naming.}}
\OA{
SymLM~\cite{jin2022symlm} extends function-name prediction by
incorporating caller, callee, and instruction-level context.
This provides additional semantic evidence when a function's role
cannot be inferred from its body alone, but depends on accurately
recovered calling context and adequate vocabulary coverage.
}

\end{itemize}

\subsection{\OA{T6: Variable Name Recovery}}

Variable-name \OA{recovery}
reconstructs meaningful identifiers for
variables whose source-level
names are unavailable after
compilation.
\OA{We group existing approaches into two categories, described below.
}

\subsubsection{Heuristic-Based Recovery}

\OA{
Heuristic-based approaches assign variable names using explicit
program patterns and known semantic cues.
}

\begin{itemize}

\item \textit{\OA{Rule- and API-Guided Naming.}}
\OA{
DREAM~\cite{yakdan2015no} and
DREAM++~\cite{yakdan2016helping}
use structural patterns and known API semantics to assign
descriptive names, including loop-related variables and arguments
passed to known functions.
For example, variables passed to \texttt{strcpy} may be named
according to their source or destination roles.
These rules provide interpretable names when known patterns are
available but generalize poorly beyond predefined cases.
}

\end{itemize}

\subsubsection{Learning-Based Recovery}

\OA{
Learning-based approaches infer variable names from statistical,
structural, and contextual representations of binary or
decompiled code.
}

\begin{itemize}

\item \textit{\OA{Statistical Alignment.}}
\OA{
Jaffe \etal~\cite{jaffe2017suggesting}
model decompiled code as a noisy translation of source code and
align decompiled variables with likely source-level identifiers.
This exploits recurring naming patterns in source-code corpora but
provides limited modeling of deeper program structure.
}

\item \textit{\OA{Feature-Based Prediction.}}
\OA{
Debin~\cite{he2018debin} combines engineered program features
with structured prediction to jointly recover variable names and
types.
This captures dependencies among program entities but remains
dependent on manually designed features and representative
training data.
}

\item \textit{\OA{Context-Aware Naming.}}
\OA{
DIRE~\cite{lacomis2019dire} incorporates lexical and AST context,
while DIRTY~\cite{chen2022augmenting} additionally uses type and
data-layout information.
ReSym~\cite{xie2024resym} and GENNM~\cite{xu2025unleashing}
further exploit broader program context and refinement mechanisms.
These richer representations improve contextual modeling but
increase complexity and remain sensitive to inaccuracies in
recovered program information.
}

\item \textit{\OA{Cross-Binary Supervised Naming.}}
\OA{
VarBERT~\cite{pal2024len} constructs supervision by compiling the
same source program into debug and stripped binaries and mapping
corresponding decompiled variables.
Debug information supplies source-level names for training,
enabling large-scale supervision without manual annotation.
The approach nevertheless depends on reliable cross-binary
mapping and representative training data.
}

\end{itemize}

\subsection{\OA{T7: Code Refinement}}\label{code_refinement}

\OA{
Code refinement transforms source-like 
decompiler output into
cleaner, more usable, or recompilable
code while aiming to
preserve program semantics.
Although recent approaches commonly rely on LLMs, 
they differ in
the primary refinement objective.
We group existing approaches into two categories, discussed next.
}

\subsubsection{Structural Refactoring}

\OA{
These approaches improve the structure and readability of
decompiled code by rewriting non-idiomatic or unnecessarily
complex constructs while preserving their intended behavior.
}

\begin{itemize}

\item \textit{\OA{Role-Based Refactoring.}}
\OA{
DeGPT~\cite{hu2024degpt} uses Referee, Advisor, and Operator
roles to determine whether refinement is needed, generate edits,
and assess their semantic consistency.
Its MSSC mechanism provides additional semantic verification,
supporting comment generation and structural simplification.
The approach avoids dedicated model training, but its
verification can be affected by complex control flow and
parsing errors.
}

\item \textit{\OA{Retrieval-Guided Refactoring.}}
\OA{
PseudoFix~\cite{li2025pseudofix} retrieves similar
distorted-pseudocode/source-code pairs as in-context examples for
refactoring decompiler output.
It targets artifacts such as non-idiomatic \texttt{goto}s,
poorly recovered loops, complex predicates, and redundant
temporaries, with additional semantic checking through
Twin-Closure.
Its effectiveness, however, depends on the relevance of retrieved
examples and remains limited for some complex transformations.
}

\end{itemize}

\subsubsection{Correctness-Oriented Refinement}

\OA{
These approaches primarily improve the syntactic, semantic, or
platform-specific correctness of decompiled code, often with the
goal of producing compilable or behaviorally consistent output.
}

\begin{itemize}

\item \textit{\OA{Feedback-Guided Repair.}}
\OA{
DecLLM~\cite{wong2025decllm} uses compiler diagnostics,
sanitizer reports, and runtime feedback to iteratively repair
syntax, type, memory, and behavioral inconsistencies.
This enables refinement toward recompilable code suitable for
source-level downstream analysis.
However, the repair process remains constrained by errors already
present in the initial decompiler output.
}

\item \textit{\OA{Architecture-Specific Refinement.}}
\OA{
ARMQwen2~\cite{liu2025armqwen2} adapts Qwen2 to ARM
decompilation and introduces preprocessing for
architecture-specific artifacts such as literal pools and
inlined data.
This specialization improves refinement of ARM-specific
decompiler output, but currently focuses on individual functions
and provides limited support for interprocedural context.
}

\end{itemize}

\subsection{\OA{T8: Code Summarization}}

\OA{
Code summarization generates natural-language descriptions of
decompiled code to recover semantic information that is not
explicitly represented in the binary.
We group existing approaches into learned and generative
summarization.
}

\subsubsection{Learned Summarization}

\OA{
Learned approaches adapt models to decompiled code through
task-specific fine-tuning or domain-specific pre-training.
BinT5~\cite{al2023extending} fine-tunes
CodeT5~\cite{wang-etal-2021-codet5} on decompiled functions
paired with source-level comments.
This adapts source-code representations to decompiler output,
but performance degrades when stripping or decompilation errors
remove relevant information.
HexT5~\cite{xiong2023hext5} instead pre-trains on pseudocode using
binary-specific objectives before applying the model to
summarization and related recovery tasks.
This captures richer binary-specific representations, but its
function-level modeling provides limited calling context.
}

\subsubsection{Generative Summarization}

\OA{
Generative approaches use general-purpose models to produce
natural-language descriptions of decompiled code through prompting.
DeGPT~\cite{hu2024degpt} uses prompt-based reasoning to determine
whether comment generation is needed and subsequently generates
semantic comments for the decompiled code.
This avoids training a dedicated summarization model, but the
generated comments may still be inaccurate and depend on the
model's interpretation of the recovered code.
}

\begin{figure*}[htbp]
    \centering
        \includegraphics[width=0.45\linewidth]{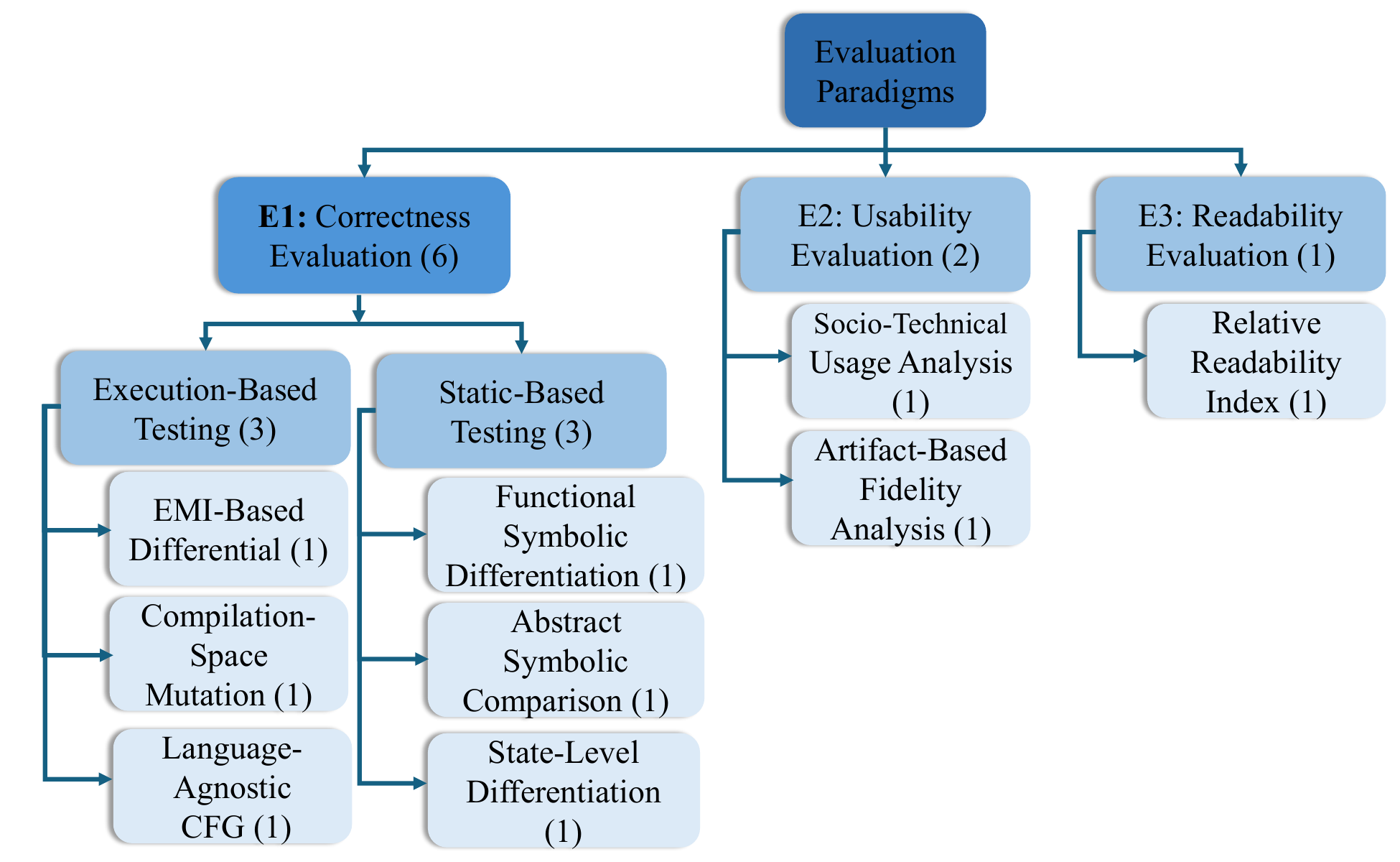}
        \caption{\OA{Taxonomy of full-pipeline decompilation evaluation
        approaches across three categories.}}
    \label{fig:evaluation_paradigms}
\end{figure*}

\section{RQ3: Research Trends in \OA{Evaluation of Decompiler Effectiveness}}\label{sec_analysisn_trends}

\OA{
In this section, we review how full-pipeline decompilers are
evaluated in terms of their effectiveness and output quality.
We organize this literature into three evaluation objectives,
as shown in~\autoref{fig:evaluation_paradigms}.
For clarity, we denote each objective as E\# and discuss them next.
}

\subsection{\OA{E1: Correctness Evaluation}}
\OA{
Correctness evaluation examines whether decompiled code
preserves the functional behavior of the original binary.
Existing studies evaluate correctness either through
execution-based testing of runtime behavior or through
static analysis of semantic equivalence.
Because some approaches require compilable or analyzable output,
decompiled code may first undergo syntactic repair or recompilation.
We group existing approaches into
two categories, described below.
}

\subsubsection{\OA{Execution-Based Testing}}

\OA{
Execution-based testing validates correctness by comparing the
runtime behavior of decompiled code with the original binary or
behaviorally equivalent program variants.
}

\begin{itemize}

\item \textit{\OA{EMI-Based Differential Testing.}}
DecFuzzer~\cite{liu2020far}
uses \textit{Equivalence Modulo Inputs (EMI)} testing to generate
program variants that are functionally equivalent for a given
input set.
By executing these variants and comparing their outputs, it
detects semantic discrepancies when behavioral divergence occurs.
\OA{
This established execution-based differential testing for
decompilers, but its evaluation is largely limited to x86 binaries
compiled at \texttt{-O0} and excludes several complex C constructs,
including structs and arrays.
}

\item \textit{\OA{Compilation-Space Mutation.}}
\OA{
Bin2Wrong~\cite{yang2025bin2wrong} extends this direction by
mutating both source constructs and compilation configurations.
It therefore exercises multiple compilers, optimization settings,
platforms, and executable formats, including ELF, PE, and Mach-O.
This broader mutation space exposes configuration-specific bugs
that EMI-based source variation may miss, at the cost of a larger
and more complex testing space.
}

\item \textit{\OA{Language-Agnostic Control-Flow Testing.}}
\OA{
FuzzFlesh~\cite{gorzynski2025fuzzflesh} moves beyond
source-language-specific testing by generating abstract CFGs and
instantiating them as Java bytecode, .NET assembly, and x86-64
machine code.
It checks whether recompiled output follows the expected runtime
path through the original CFG, enabling cross-platform detection
of control-flow misdecompilation.
This improves portability across execution environments, but
focuses primarily on control-flow correctness rather than the full
semantics of decompiled programs.
}

\end{itemize}

\subsubsection{\OA{Static-Based Testing}}

\OA{
Static-based testing compares semantic representations of the
original binary and decompiled output without relying solely on
concrete executions.
}

\begin{itemize}

\item \textit{\OA{Functional Symbolic Differentiation.}}
D-Helix~\cite{zou2024d} includes a
\texttt{RECOMPILER} component that repairs and recompiles
decompiler output before semantic comparison.
It then uses \texttt{SYMDIFF} to compare symbolic models of the
original lifted binary IR and the recompiled output based on
functional equivalence of return semantics.
D-Helix further introduces \texttt{TUNER} to isolate faulty
decompiler heuristics by selectively toggling them during
recompilation.
This enables automated root-cause analysis, but return-oriented
equivalence may overlook intermediate side effects and requires
successful recompilation.

\item \textit{\OA{Abstract Symbolic Comparison.}}
Cao \etal~\cite{cao2024evaluating} avoid recompilation by
abstracting optimized LLVM IR and decompiled code into comparable
symbolic representations.
Using \textit{Dsmith}-generated programs with embedded checksums,
the framework compares control-flow markers and path-level data
semantics through concolic testing.
\OA{
Although they also evaluate
TransRepair~\cite{li2022transrepair} for syntactic repair,
their comparison approach does not depend on it.
}
\OA{
This enables evaluation without successful recompilation,
but requires specialized test generation and symbolic traceability.
}

\item \textit{\OA{State-Level Differentiation.}}
\OA{
DiscScope~\cite{sirlanci2025empirical} extends comparison from
end-point behavior to intermediate program states by symbolically
executing the original and recompiled binaries in parallel.
It localizes divergences through variables, stack offsets,
literals, control-flow regions, and statements, while
resynchronizing executions when their control flow diverges.
This provides finer-grained diagnosis than return-level
comparison, but remains dependent on recompilable decompiler
output.
}

\end{itemize}

\subsection{\OA{E2: Usability Evaluation}}
\OA{
Usability evaluation examines how decompiler outputs and tools
support reverse-engineering practice.
Existing studies evaluate usability either through
socio-technical analysis of analyst interactions or through
human-centered examination of decompiled-code artifacts.
We group existing approaches into 
three categories, described below.
}

\subsubsection{\OA{Socio-Technical Usage Analysis}}
Votipka \etal~\cite{votipka2021investigation}
combine \OA{qualitative coding, statistical analysis, and social
network analysis to examine community engagement} with
Ghidra~\cite{ghidra} after its public release.
They collected 1,590 discussion threads from 
StackExchange~\cite{StackExchange}, Reddit~\cite{reddit}, and 
Twitter~\cite{twitter}, covering 688 users over six months.
A sample of the threads was qualitatively coded along dimensions 
such as discussed features, conversational structure, and forms of 
sensemaking.
Their findings show that Ghidra's decompiler became a central focus 
of discussion, both as a practical reverse-engineering tool and as 
a target for extension through scripting.
The study also shows that community support often followed a 
producer-consumer pattern, where a smaller group of experienced 
users answered questions from a larger user base.
\OA{
This provides evidence of real-world usability and adoption
patterns, but relies on public discourse rather than directly
evaluating decompiled-code artifacts.
}

\subsubsection{\OA{Artifact-Based Fidelity Analysis}}
\OA{
Dramko \etal~\cite{dramko2024taxonomy}
compare original and decompiled C code from Hex-Rays, Ghidra,
RetDec, and angr using open coding, thematic analysis, and
inter-coder agreement.
}
This process yielded a hierarchical taxonomy comprising 15 
top-level categories and 52 subcodes, covering both correctness 
issues (\eg incorrect return behavior and type-recovery errors) 
and readability concerns (\eg redundant expressions and missing 
type hints).
The study further distinguishes between issues that are more 
deterministic, and may be addressed through heuristics or more 
accurate type information, and issues that are more context 
dependent, requiring external information such as macros, 
identifier names, or learned models.

\subsection{\OA{E3: Readability Evaluation}}

\OA{
Decompiler output may preserve program behavior while remaining
difficult for analysts to understand due to structural and
syntactic artifacts.
Readability evaluation captures this human-facing dimension beyond
functional correctness.
}

\subsubsection{\OA{Relative Readability Index}}
Eom \etal~\cite{eom2024r2i}
introduced R2I, a readability 
metric for decompiler output
that produces a normalized 
score between 0 and 1.
\OA{
R2I first applies rule-based repairs to malformed
decompiler output to make it parsable.
}
It \OA{then} 
extracts 31 AST-based static 
features capturing syntactic
structure and common decompilation 
artifacts, with feature weights
partly informed by reverse-engineer surveys.
\OA{
The metric enables relative 
comparison across decompilers,
versions, and obfuscation levels and 
correlates with human
readability judgments.
}
\OA{
However, its weighting reflects surveyed 
user preferences and may
therefore vary with analyst 
expertise and downstream tasks.
}

\begin{table*}[t!]
    \centering
    \caption{
        \OA{
            Summary of evaluation capabilities reported across decompilation studies.
        }
       }
    \resizebox{0.75\linewidth}{!}{%

\begin{tabular}{p{2.6cm}p{5.2cm}p{8.2cm}}
\toprule
\textbf{Capability} & \textbf{Representative Metrics} & \textbf{Representative Papers} \\
\midrule

Compactness &
Compression ratio 
(\eg $1 - \frac{\mathrm{LOC}_{dec}}{\mathrm{LOC}_{asm}}$);
reduction ratio 
(\eg percentage reduction in code volume);
LOC reduction;
number of expressions;
temporary variables
&
DCC~\cite{dcc,cifuentes1995decompilation},
C-Decompiler~\cite{c_decompiler_conf,chen2013refined},
REcompile~\cite{yakdan2013recompile},
DREAM~\cite{yakdan2015no},
ERASE~\cite{zhang2024optimizing},
PseudoFix~\cite{li2025pseudofix} \\

\midrule

Structuredness &
Goto count 
(\eg number of emitted \texttt{goto}s);
goto elimination;
structured-region coverage;
control-flow recovery;
cyclomatic complexity 
(\eg number of independent paths);
nesting depth;
control-flow complexity;
CFG edit distance &
Phoenix~\cite{brumley2013native},
C-Decompiler~\cite{c_decompiler_conf,chen2013refined},
ERASE~\cite{zhang2024optimizing},
PseudoFix~\cite{li2025pseudofix},
SAILR~\cite{basque2024ahoy} \\
\midrule

Readability &
Cyclomatic complexity;
Halstead effort;
static code features;
LOC, variables, casts, branches, and gotos;
variable expansion rate 
(\eg variables in decompiled code relative to source variables);
human assessment;
user preference ranking;
LLM rubric score &
C-Decompiler~\cite{c_decompiler_conf,chen2013refined},
REcompile~\cite{yakdan2013recompile},
BED~\cite{schulte2018evolving},
rev.ng COMB~\cite{gussoni2020comb},
dewolf~\cite{enders2022dewolf},
DeGPT~\cite{hu2024degpt},
ERASE~\cite{zhang2024optimizing},
LLM4Decompile~\cite{tan2024llm4decompile},
SAILR~\cite{basque2024ahoy} \\

\midrule

Correctness &
Test-suite execution;
re-executability;
functional matching;
basic-block semantic equivalence;
input-output accuracy;
differential execution checksum;
MSSC;
online judge validation;
byte similarity
&
Phoenix~\cite{brumley2013native},
DREAM~\cite{yakdan2015no},
BED~\cite{schulte2018evolving},
FoxDec~\cite{verbeek2020sound},
NeurDP~\cite{cao2022boosting},
DeGPT~\cite{hu2024degpt},
ERASE~\cite{zhang2024optimizing},
LLM4Decompile~\cite{tan2024llm4decompile},
SLaDe~\cite{armengol2024slade}
\\
\midrule

Completeness &
Recovered functions;
callee/caller recovery;
CFG edge coverage;
feasible control-transfer coverage;
type inference coverage;
recovery rate &
Phoenix~\cite{brumley2013native},
angr~\cite{shoshitaishvili2016sok},
TYGR~\cite{zhu2024tygr} \\
\midrule

Usability &
User studies;
task success rate 
(\eg correctly solved analysis tasks);
analyst effort;
task completion time;
number of errors;
Likert-scale feedback;
ratings;
correctness of human understanding &
DREAM++~\cite{yakdan2016helping},
Decomperson~\cite{burk2022decomperson},
ERASE~\cite{zhang2024optimizing},
VarBERT~\cite{pal2024len},
PseudoFix~\cite{li2025pseudofix}
\\

\midrule

Similarity &
Classification Metrics (\eg P, R, F1);
Edit distance 
(\eg token- or character-level edits);
average edit distance;
character error rate;
BLEU;
ROUGE-L;
METEOR;
token match;
embedding-based semantic similarity &
Katz \etal~\cite{katz2018using},
Coda~\cite{fu2019coda},
BTC~\cite{hosseini2022beyond},
BinT5~\cite{al2023extending},
HexT5~\cite{xiong2023hext5},
DIRE~\cite{lacomis2019dire},
PseudoFix~\cite{li2025pseudofix} \\

\midrule

Recompilability &
Compilation success rate 
(\eg successfully compiled outputs over all outputs);
multi-configuration compilation;
repair success rate &
FoxDec~\cite{verbeek2020sound},
LLM4Decompile~\cite{tan2024llm4decompile},
ARMQwen2~\cite{liu2025armqwen2},
DecLLM~\cite{wong2025decllm} \\

\bottomrule
\end{tabular}
    }
    \label{tbl:metric_dec}
\end{table*}

\section{RQ4: Research Trends in \OA{Evaluating Decompilation Capabilities}}
\label{sec_mtrics_trends}

\OA{
In this section, we analyze how decompilation studies evaluate
eight decompiler capabilities, summarized in
\autoref{tbl:metric_dec}.
For each capability, we ask three questions about
what it means, how it is quantified, and what it implies for
decompiled code and reverse-engineering workflows.
}
Developing an approach that simultaneously
satisfies multiple \OA{capabilities}
is inherently challenging, as they
often capture orthogonal aspects of quality, 
\OA{discussed next.}

\subsection{Compactness}

\OA{Compactness refers to how much the decompiled output reduces 
low-level code volume and redundancy.}
\OA{For reverse engineers, compact code can reduce the amount of 
text, expressions, and temporary variables that must be inspected.
} 

\begin{itemize}
    \item \textit{\OA{Code-Volume Reduction.}}
    Dcc~\cite{cifuentes1995decompilation}, 
    C-Decompiler~\cite{c_decompiler_conf,chen2013refined}, 
    REcompile~\cite{yakdan2013recompile}, and 
    DREAM~\cite{yakdan2015no} measure compactness by comparing 
    the size of the decompiled code with its assembly or source 
    counterpart, often through a normalized compression ratio such as
    \( 1 - \frac{\text{LOC}_{\text{decompiled}}}{\text{LOC}_{\text{assembly}}} \).
    \OA{Later work \OA{extends this measurement to} related 
     indicators,
    including expression count and temporary-variable count
    ~\cite{zhang2024optimizing,li2025pseudofix}.
    }
\end{itemize}

\subsection{Structuredness}

Structuredness refers to how well 
decompiled output recovers 
high-level control-flow constructs, 
such as conditionals and loops, 
instead of exposing low-level jumps.
\OA{For reverse engineers, 
structured code reduces the need to 
manually trace arbitrary branches
and makes program logic easier 
to follow.
}

\begin{itemize}
    \item \textit{\OA{Goto Count.}}
    Phoenix~\cite{brumley2013native} evaluates
    structuredness by counting the number of
    \texttt{goto} statements in the output.
    These statements often indicate that the decompiler
    could not recover structured control flow from the
    underlying CFG.
    Since excessive \texttt{goto} usage is associated with
    difficult-to-maintain ``spaghetti code''~\cite{dijkstra1968letters},
    fewer \texttt{goto} statements generally indicate better
    structural recovery.
    
    \item \textit{\OA{Control-Flow Structure.}}
    \OA{
    Beyond counting unstructured jumps,
    structuredness can also be evaluated by comparing
    the recovered control-flow structure with the
    source-level structure.
    SAILR~\cite{basque2024ahoy} uses
    CFG Edit Distance (CFGED) to quantify
    structural deviations between them.
    Lower edit distance indicates greater structural
    similarity to the source-level program.
    Other studies use related measures,
    including structured-region coverage,
    cyclomatic complexity, nesting depth, and
    control-flow similarity
    ~\cite{c_decompiler_conf,chen2013refined,
    zhang2024optimizing,li2025pseudofix,basque2024ahoy}.
    }
\end{itemize}

\subsection{Readability}

\OA{
Readability refers to how easily decompiled code can be
understood as high-level source code.
For reverse engineers, readable output reduces the effort
needed to infer program intent and follow control and data relationships.
Because some aspects of readability may be perceived subjectively,
studies combine static code metrics with human-centered assessments.
}

\begin{itemize}
    \item \textit{Structural Complexity.}
    C-Decompiler~\cite{c_decompiler_conf,chen2013refined} and
    rev.ng COMB~\cite{gussoni2020comb}
    assess readability using control-flow characteristics, 
    especially cyclomatic complexity~\cite{mccabe1976complexity}, 
    which captures the number of independent execution paths.
    \OA{Lower complexity generally indicates code that is easier to 
    follow, although it does not fully capture semantic clarity.}

    \item \textit{Static Code Features.}
    BED~\cite{schulte2018evolving} uses static code features, 
    such as the number of characters, casts, variable declarations, 
    and control-flow constructs, as indirect indicators of 
    cognitive load.
    SAILR~\cite{basque2024ahoy} similarly highlights lines of code, 
    Boolean expressions, and \texttt{goto} statements as readability 
    indicators.

    \item \textit{Lexical Complexity.}
    DeGPT~\cite{hu2024degpt} applies 
    Halstead's complexity~\cite{halstead1977elements}, 
    which estimates cognitive effort from the number of distinct 
    operators and operands in the recovered code.
    \OA{This captures how dense or verbose the output is at the 
    token level.}

    \item \textit{Survey and User Assessment.}
    REcompile~\cite{yakdan2013recompile} and 
    dewolf~\cite{enders2022dewolf} evaluate readability through 
    human judgments collected from questionnaires, interviews, or 
    controlled user studies.
    \OA{These assessments capture perceived clarity, but their 
    results may depend on participant expertise and task context.}

    \item \textit{Hybrid Static and Human Evaluation.}
    ERASE~\cite{zhang2024optimizing} combines human evaluation with 
    static analysis.
    Participants inspect decompiled functions, while the system 
    also measures properties such as lines of code, variable count, 
    branch count, control-flow depth, and \texttt{goto} usage.

    \item \textit{Rubric-Based LLM Evaluation.}
    LLM4Decompile~\cite{tan2024llm4decompile} uses GPT-4o as an 
    evaluator with a structured rubric for syntax and structural 
    similarity between decompiled output and source code.
    The score ranges from 1 (poor) to 5 (excellent), providing a 
    systematic but model-dependent readability assessment.
\end{itemize}

\subsection{Correctness}

\OA{Correctness refers to whether the decompiled output preserves 
the behavior of the original binary.}
\OA{For reverse engineers, correct output is essential because 
analysis performed on incorrect decompiled code may lead to false 
conclusions about program behavior.}

\begin{itemize}

\item \textit{Execution-Based Validation.}
DREAM~\cite{yakdan2015no}, Phoenix~\cite{brumley2013native} and \OA{LLM4Decompile~\cite{tan2024llm4decompile}} 
assess correctness by inserting recovered functions back into 
the original source code and running high-coverage test suites.
Successful execution indicates that semantic behavior is preserved.

\item \textit{Basic-Block Semantic Equivalence.}
NeurDP~\cite{cao2022boosting} measures correctness by comparing 
the semantics of corresponding basic blocks in the original 
low-level representation and the decompiled high-level 
representation.
Each basic block is treated as a mapping from inputs to outputs, 
and correctness is determined by whether both representations 
produce matching outputs for the same inputs.

\item \textit{Simulation-Based Semantic Checking.}
DeGPT~\cite{hu2024degpt} applies Micro Snippet Semantic 
Calculation (MSSC), which simulates function logic to verify 
return values and side effects without requiring full program 
execution.
\OA{This provides a lightweight correctness signal when complete 
recompilation or end-to-end execution is difficult.}

\item \textit{Online Judge System.} 
ERASE~\cite{zhang2024optimizing} evaluates correctness by 
submitting decompiled code
to the LeetCode online judge system~\cite{LeetCode}, 
using automated I/O test case verification.

\end{itemize}

\subsection{Completeness}

\OA{Completeness refers to the extent to 
which decompilation 
recovers the program elements needed 
to represent the original 
binary.}
\OA{For reverse engineers, incomplete 
recovery can hide relevant 
functions, control-flow paths, or 
type information, leading to 
partial or misleading 
program understanding.}

\begin{itemize}

\item \textit{Function Recovery.}
Phoenix~\cite{brumley2013native} demonstrates that 
if a callee function \(g\) is not recovered, 
any caller function \(f\) that invokes
it may also fail to decompile.
This dependency illustrates 
how missing components can
cascade and affect the
recovery of related code.

\item \textit{CFG \OA{Recovery}.}
angr~\cite{shoshitaishvili2016sok} \OA{captures completeness 
through CFG recovery, where the goal is to cover feasible control 
transfers in the original binary.}
\OA{However, broader CFG coverage may require over-approximation, 
which can introduce spurious edges and trade precision for 
completeness.}

\item \textit{\OA{Type Recovery}.}
TYGR~\cite{zhu2024tygr} \OA{measures type-coverage limitations by 
tracking how many observed variable types in real-world binaries 
can be represented and inferred using its fixed type vocabulary.}
Types outside this vocabulary are 
approximated using fallback casts
(\eg \texttt{struct***} $\rightarrow$ \texttt{void*}).

\end{itemize}

\subsection{Usability}

\OA{
Usability refers to how effectively decompiled output supports
reverse-engineering tasks.
For reverse engineers, usable output should reduce analysis effort,
improve task accuracy, and increase confidence in the recovered
code.
}

\begin{itemize}

\item \textit{Task-Based Performance.}
DREAM++~\cite{yakdan2016helping} evaluates usability by asking
participants to complete reverse-engineering tasks and measuring
task success rate, completion time, and errors.
\OA{
ERASE~\cite{zhang2024optimizing} similarly measures analysis time
before and after modularity restoration to quantify changes in
manual effort.
These measures capture how effectively decompiled output supports
analysts in completing practical reverse-engineering tasks.
}

\item \textit{Subjective Feedback.}
\OA{
DREAM++~\cite{yakdan2016helping} and
Decomperson~\cite{burk2022decomperson} use participant feedback to
assess perceived readability, clarity, and usefulness.
Such measures capture user perception but may vary with participant
expertise and task context.
}

\item \textit{\OA{Perceived Structural Quality.}}
\OA{
PseudoFix~\cite{li2025pseudofix} evaluates perceived structural
quality through human judgments of consistency and reasonability.
Consistency assesses whether transformed pseudocode follows
structural patterns similar to a reference sample, whereas
reasonability assesses whether the code appears logically coherent
and well structured.
}

\end{itemize}

\subsection{\OA{Similarity}}
\OA{
Similarity measures how closely decompiler output matches a
ground-truth source representation, identifier, summary, or code fragment.
For reverse engineers, it provides a convenient proxy for output quality,
but does not necessarily imply semantic correctness or readability.
Different decompilation tasks therefore adopt different similarity measures.
}

\begin{itemize}

\item \textit{\OA{Classification Metrics.}}
\OA{
Precision, recall, and F1 score measure agreement with
ground-truth labels while accounting for false positives and
false negatives.
They are commonly used in recovery tasks such as type inference,
as in OSPREY~\cite{zhang2021osprey}.
}

\item \textit{\OA{Exact Match and Accuracy.}}
\OA{
Exact match and accuracy measure whether a predicted label,
identifier, or code fragment matches the reference exactly.
They are used in tasks such as variable name
recovery~\cite{pal2024len}.
}

\item \textit{\OA{Edit Distance and Character Error Rate.}}
\OA{
These metrics quantify the token- or character-level changes
required to transform generated output into the reference.
They are used for source-code generation and identifier recovery
~\cite{katz2018using,hosseini2022beyond,lacomis2019dire}.
}

\item \textit{\OA{N-Gram-Based Similarity.}}
\OA{
BLEU, ROUGE-L, and METEOR measure sequence overlap between
generated and reference outputs.
They are commonly used for code generation and summarization
~\cite{al2023extending,xiong2023hext5,tan2025decompile,
li2025pseudofix}.
}

\item \textit{\OA{Embedding-Based Code Similarity.}}
\OA{
Embedding-based metrics compare semantic representations beyond
surface-level token overlap.
For example, PseudoFix~\cite{li2025pseudofix} uses
CodeBERT-based similarity to compare refactored pseudocode with
reference code.
}

\end{itemize}

\subsection{Recompilability}
\OA{Recompilability refers to whether decompiled code can be 
accepted by a standard compiler without manual repair.}
\OA{For reverse engineers, recompilable output enables downstream 
uses such as testing, patching, instrumentation, and program 
analysis on recovered source code.}
\begin{itemize}

\item \textit{Compilation Success.}
FoxDec~\cite{verbeek2020sound} 
evaluates recompilability by
compiling the recovered C code 
under different compiler
configurations
(\eg compiler types, versions, and optimization levels).
\OA{
This treats recompilability as a direct pass/fail outcome, where
the generated code either compiles successfully or does not.
LLM4Decompile~\cite{tan2024llm4decompile},
ARMQwen2~\cite{liu2025armqwen2}, and
DecLLM~\cite{wong2025decllm}
similarly report compilation success after refinement processes.
}

\end{itemize}

\section{RQ5: Research Trends 
in Implementation and Benchmarks}
\label{sec_implementation_trends}

In this section, we examine the infrastructure
supporting decompilation research, including
implementation setups, analysis \OA{toolchains},
ML architectures, benchmark construction, and
compiler configurations.
\OA{These elements characterize how
decompilation systems are implemented and
evaluated in practice.
}

\begin{figure*}[htbp]
    \centering
        \includegraphics[width=0.99\linewidth]{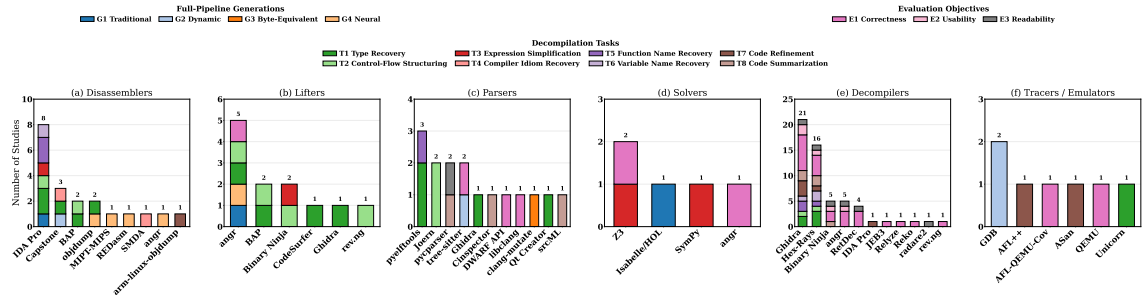}
        \caption{
             Distribution of analysis tools, \OA{
             used across the reviewed decompilation studies.
             Colors indicate their use 
             across full-pipeline generations (G\#), 
             decompilation tasks (T\#), and 
             evaluation objectives (E\#).
             }
       }
    \label{fig:tool_popularity}
\end{figure*}

\subsection{\OA{Analysis Toolchain}}

\OA{
Decompilation research relies on a diverse analysis toolchain to
support different stages of the workflow.
~\autoref{fig:tool_popularity} shows these tools range from
static components to dynamic components.
We next examine their roles, 
adoption across decompilation research,
and key design trade-offs.
}

\subsubsection{Disassemblers}

Disassemblers convert binary code into assembly and serve as the
entry point for most decompilation pipelines.
\OA{
Errors in instruction boundaries or control transfers can
propagate to downstream analyses~\cite{pang2021sok}.
}
\autoref{fig:tool_popularity}(a) \OA{
shows their use across G1, G2, and G4 and seven task-oriented
techniques, with T1 being the most frequent.
IDA Pro~\cite{ida_pro} dominates usage because of its mature
interactive analysis and scripting support, while Capstone
~\cite{capstone} provides a lightweight embeddable alternative
and BAP~\cite{bap} offers broader program-analysis support.
}
\OA{
These choices reflect different needs for interactive analysis,
lightweight integration, and richer downstream reasoning.
}

\subsubsection{Binary Lifters}

Binary lifters translate assembly instructions into intermediate
representations, enabling architecture-agnostic analyses such as
data-flow tracking, type recovery, and symbolic reasoning.
\OA{
Their semantic accuracy is important because incomplete or
incorrect instruction lifting can propagate errors to subsequent
decompilation analyses~\cite{liu2022sok}.
}
\autoref{fig:tool_popularity}(b)
\OA{
shows that binary lifters are used across
G1 and G4, four task-oriented techniques, and E1.
They are used most frequently for T1 and T2,
with four studies each.
}
\OA{
angr~\cite{angr} dominates usage, appearing in five studies,
due to its VEX-based IR, Python interface, and program-analysis
support.
BAP~\cite{bap}, Binary Ninja, CodeSurfer~\cite{codesurfer_x86},
rev.ng~\cite{revng}, and Ghidra provide alternatives with
different IRs, architecture coverage, and analysis interfaces.
}
\OA{
Overall, lifter selection reflects a trade-off among semantic
precision, IR expressiveness, architecture coverage, and
downstream integration.
}

\subsubsection{Parsers}

Parsers extract program structure, syntax, or metadata from
binaries, source code, and decompiled output for analysis,
training, and evaluation.
\autoref{fig:tool_popularity}(c) \OA{
shows their use across 15 studies, with T1 and E1 being the most
frequent applications.
pyelftools~\cite{pyelftools} is the most common choice for binary
metadata, while Joern~\cite{joernio_joern},
pycparser~\cite{EliBendersky_pycparser}, and
tree-sitter~\cite{tree_sitter} provide graph-, AST-, and
syntax-oriented representations.
}
\OA{
Parser selection therefore depends primarily on whether the
analysis requires binary metadata, source syntax, or richer
structural representations.
}

\subsubsection{\OA{Solvers}}

Solvers support symbolic reasoning over expressions, constraints,
and program states that cannot be resolved syntactically.
\autoref{fig:tool_popularity}(d) \OA{
shows their use in G1, T3, and E1.
Z3~\cite{Z3} dominates, while SymPy~\cite{SymPy},
angr~\cite{angr}, and Isabelle/HOL~\cite{isabelle} represent
algebraic simplification, symbolic execution, and formal
verification, respectively.
}
\OA{
Their use trades greater semantic reasoning power for increased
analysis complexity and computational cost.
}

\subsubsection{Decompilers}

Decompilers are used as experimental subjects, baselines, or
sources of decompiled code for downstream analysis and refinement.
\autoref{fig:tool_popularity}(e) \OA{
shows that they form the largest infrastructure category,
spanning six task-oriented techniques and all three evaluation
objectives.
Ghidra~\cite{ghidra} and Hex-Rays~\cite{hexrays_decompiler}
dominate usage, reflecting their mature pipelines and frequent
use in comparative evaluation and downstream processing.
}
\OA{
Other decompilers broaden tool diversity for robustness and
cross-decompiler comparison.
}

\subsubsection{Tracers \& Emulators}

Although decompilation is primarily static, some studies use
dynamic execution to obtain runtime evidence unavailable through
static analysis alone.
\autoref{fig:tool_popularity}(f) \OA{
shows five studies spanning G2, T1, T7, and E1.
GDB~\cite{gdb_manual} is the most frequent, while emulation,
instrumentation, and fuzzing tools provide complementary runtime
support.
}
\OA{
Their use trades additional execution cost for direct observation
of program behavior.
}

\begin{figure*}[htbp]
    \centering
        \includegraphics[width=0.8\linewidth]{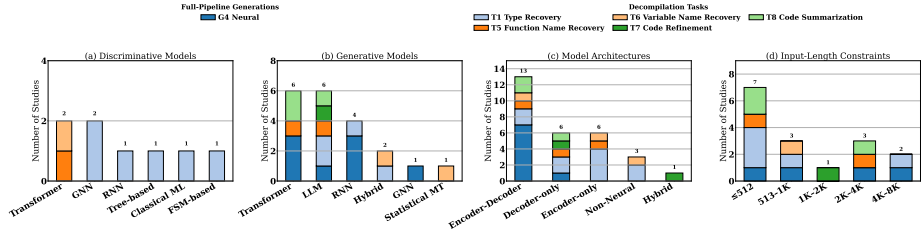}
        \caption{
            Distribution of ML models by \OA{learning objectives} (\eg discriminative vs. generative), architecture, and token limits.
        }
    \label{fig:ml_family}
\end{figure*}

\subsection{\OA{Model Choices}}
\label{sec:model_choices}

\OA{
To understand how learning is incorporated into modern
decompilation, we examine what models predict, which model
families they adopt, how they are architecturally designed,
and how much input context they can process, as summarized in
\autoref{fig:ml_family}.
}

\subsubsection{\OA{Learning Objectives}}

\OA{
Learning-based approaches primarily follow discriminative or
generative objectives, depending on whether they predict
predefined outputs or generate new ones.
}

\begin{itemize}

\item \textit{\OA{Discriminative Models.}}
Discriminative models classify or label elements of binary or
decompiled code, making them suitable for constrained prediction
tasks.
As shown in \autoref{fig:ml_family}(a), \OA{
eight surveyed models follow this objective.
Six target T1, while T5 and T6 each account for one model.
Transformers and GNNs are the most frequent model families,
with two models each, while recurrent, tree-based, finite-state,
and classical machine-learning models each appear once.
}
\OA{
These choices support different forms of program context, but
their predefined output spaces make them less flexible for
open-ended reconstruction.
}

\item \textit{\OA{Generative Models.}}
Generative models synthesize sequences or structured
representations from binary, assembly, or decompiled-code inputs.
As shown in \autoref{fig:ml_family}(b), \OA{
20 surveyed models follow this objective.
Eight target G4, followed by four targeting T1, three T8,
two each T5 and T6, and one T7.
Transformers and LLMs dominate with six models each, followed
by recurrent models with four and hybrid designs with two;
GNN and statistical machine translation each appear once.
}
\OA{
Generative models provide greater flexibility for open-ended
reconstruction, but their larger output space can introduce
syntactic or semantic errors.
}

\end{itemize}

\subsubsection{\OA{Model Architecture}}
\OA{
\autoref{fig:ml_family}(c) shows that encoder--decoder models
dominate, with 13 studies, followed by encoder-only and
decoder-only architectures with six each.
Encoder--decoder designs remain common for sequence-to-sequence
translation, encoder-only models support constrained prediction,
and decoder-only architectures reflect the recent adoption of
GPT-style LLMs for open-ended generation.
Non-neural and hybrid designs are comparatively rare.
}

\subsubsection{\OA{Input-Length Constraints}}
\label{sec:model_context_length}

Token length determines the maximum amount of tokenized input
that a model can process in a single invocation.
\OA{
This constraint is important in decompilation because low-level
representations and long functions can expand substantially
after tokenization.
}
As shown in \autoref{fig:ml_family}(d), \OA{
reported limits are concentrated in five ranges.
Seven models operate at $\leq512$ tokens, three between
513 and 1K, one between 1K and 2K, three between 2K and 4K,
and two between 4K and 8K.
}
\OA{
For example, BinT5~\cite{al2023extending} and
HexT5~\cite{xiong2023hext5} use 512-token limits, whereas
LLM4Decompile~\cite{tan2024llm4decompile} and
DeGPT~\cite{hu2024degpt} use 4,096 tokens.
Larger configurations include
BTC~\cite{hosseini2022beyond} with 6,350 tokens and
ReSym~\cite{xie2024resym} with 8,192.
}
\OA{
These values represent study-specific configurations rather
than fixed limits of a model family.
Context capacity can vary across model versions
(\eg successive GPT-family models), while studies may also
adopt smaller limits because of memory, batching, fine-tuning,
or dataset constraints.
}

\begin{wrapfigure}{r}{0.52\columnwidth}
    \centering
    \includegraphics[width=\linewidth]
    {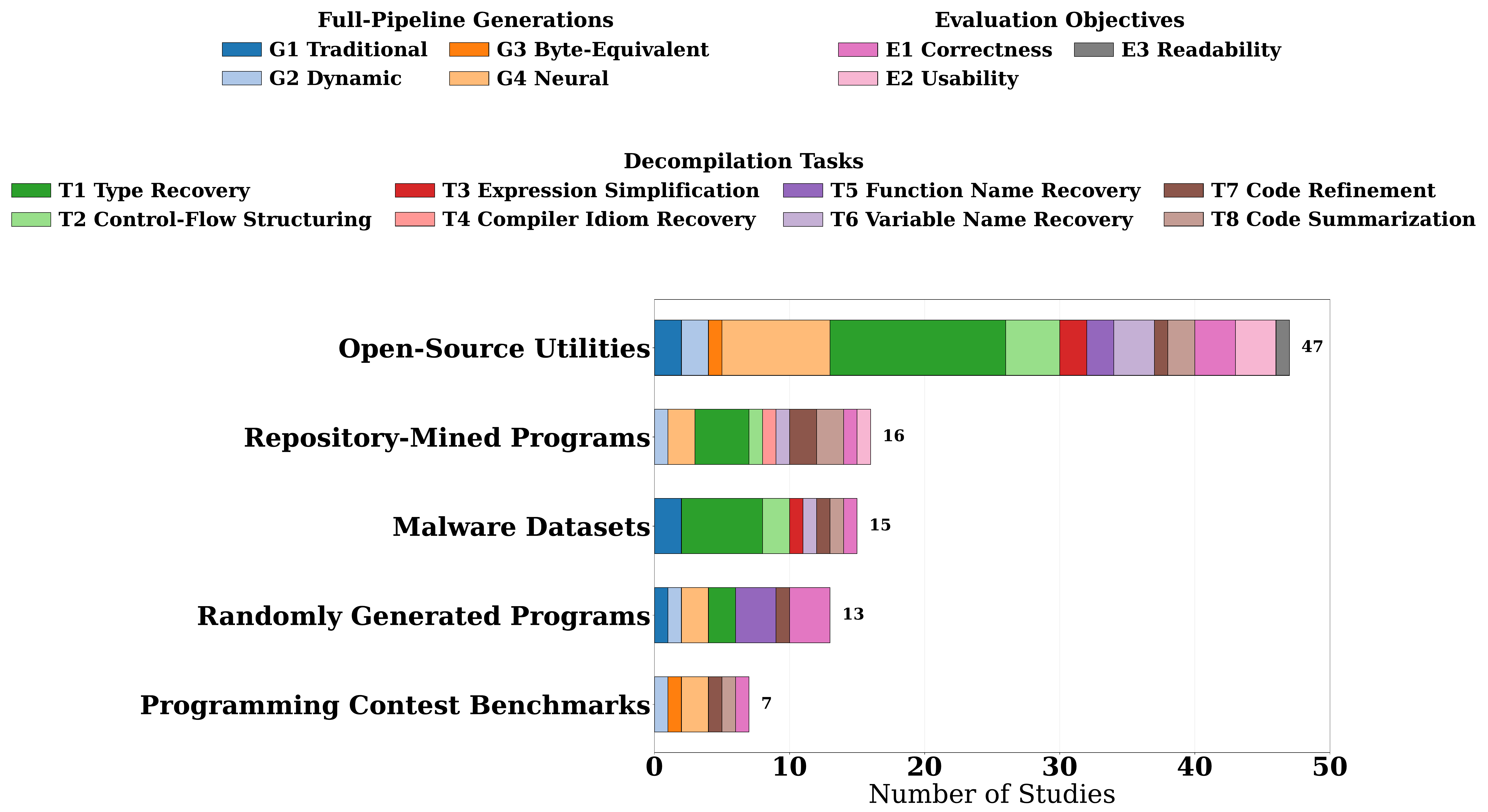}
    \caption{\OA{
        Distribution of benchmark suites.
    }}
    \label{fig:benchmark}
\end{wrapfigure}

\subsection{Benchmark Suites}  

To evaluate the effectiveness and  
generalizability of  
decompilation techniques,  
researchers rely on diverse  
benchmark datasets.  
These datasets are selected to  
reflect real-world conditions,  
test specific capabilities,  
or introduce challenging scenarios.
\OA{
We group these datasets into
five categories, shown in~\autoref{fig:benchmark}
and discussed next.
}

\subsubsection{Open-Source Utilities}

Open-source utilities provide real-world programs with
accessible source code and reproducible build environments.
\OA{
Their availability makes them suitable for repeatable
decompilation evaluation and comparison across studies.
}
\autoref{fig:benchmark}
\OA{
shows that open-source utilities are the most widely used
benchmark family.
Other utility programs appear in 37 studies, primarily in
T1 and G4, with ten and eight studies, respectively.
GNU Coreutils~\cite{gnu2025coreutils} appears in 24 studies,
with its use concentrated in T1, with 11 studies, followed
by T2, with three.
}
\OA{
Other utilities, including glibc~\cite{glibc},
Karel~\cite{pattis1981karel}, and
Hacker's Delight~\cite{warren2002hackers},
provide alternative program characteristics.
Overall, open-source utilities offer strong reproducibility
and practical relevance, but repeated reliance on a limited
set of software families may reduce benchmark diversity.
}

\subsubsection{Large-Scale Repository Mining}

Large-scale repository mining constructs benchmark corpora from
public repositories or large software distributions.
\OA{
Its scale is particularly important for learning-based
decompilation, where large numbers of binary--source pairs are
required for training and evaluation.
}
\autoref{fig:benchmark}
\OA{
shows that repository-mined programs appear in 16 studies,
spanning G2 and G4, six task-oriented techniques, and E1--E2.
They are used most frequently for T1, with four studies,
followed by G4, T7, and T8, with two studies each.
}
\OA{
These corpora are commonly obtained from repository collections
such as GHTorrent~\cite{gousios2013ghtorent} and GHCC~\cite{ghcc},
or software distributions such as
Debian~\cite{DebianBuster} and Fedora~\cite{FedoraRPM}.
Repository mining provides scale and software diversity, but
differences in collection and filtering procedures limit
cross-study comparability.
}

\subsubsection{Malware Datasets}

Malware datasets expose decompilers to security-relevant and
potentially irregular binary behavior.
\OA{
They complement conventional software benchmarks by providing
evaluation settings closer to practical malware analysis.
}
\autoref{fig:benchmark}
\OA{
shows that malware datasets appear in 15 studies, spanning
G1, six task-oriented techniques, and E1.
Their use is concentrated in T1, with six studies, followed
by G1 and T2, with two studies each.
}
\OA{
Recurring samples include SpyEye, Cridex, ZeusP2P, and Mirai,
while other malware families appear more selectively.
Malware datasets improve realism for security-oriented
evaluation, but limited availability and inconsistent sample
sharing hinder reproducibility.
}

\subsubsection{Random Program Generators}

Random program generators synthesize valid C programs for
controlled and systematic decompiler testing.
\OA{
Their ability to generate diverse inputs makes them useful for
stress testing and correctness-oriented evaluation.
}
\autoref{fig:benchmark}
\OA{
shows that randomly generated programs appear in 13 studies,
spanning G1, G2, G4, three task-oriented techniques, and E1.
They are used most frequently for E1 and T5, with three studies
each, followed by G4 and T1, with two each.
}
\OA{
Csmith~\cite{csmith_project} is the dominant generator,
while Dsmith~\cite{cao2024evaluating} and
Cfile~\cite{cfile_project} provide alternative generation
strategies.
}
Though useful for stress testing, these generators
are seldom used and may poorly reflect
the complexity of real-world code.

\subsubsection{Programming Contest Datasets}

Programming-contest datasets provide compact, standardized
programs with limited external dependencies.
\OA{
These properties simplify dataset construction and make them
particularly suitable for learning-based decompilation.
}
\autoref{fig:benchmark}
\OA{
shows that programming-contest datasets appear in seven studies,
spanning G2--G4, T7--T8, and E1.
G4 accounts for the largest share, with two studies, while each
remaining category appears once.
}
\OA{
LeetCode~\cite{LeetCode} is the principal source, while
Project Euler~\cite{ProjectEuler} provides primarily
arithmetic-oriented programs.
These datasets facilitate controlled comparison, but provide
limited coverage of real-world dependencies and software
complexity.
}

\OA{
Beyond these benchmark categories,
Decompile-Bench~\cite{tan2025decompile}
represents a recent effort toward standardized benchmark
construction using large-scale binary--source pairs from
real-world software repositories.
It therefore complements existing benchmark sources with a
more systematic and reproducible construction methodology.
}

\begin{figure*}[htbp]
    \centering
        \includegraphics[width=0.99\linewidth]{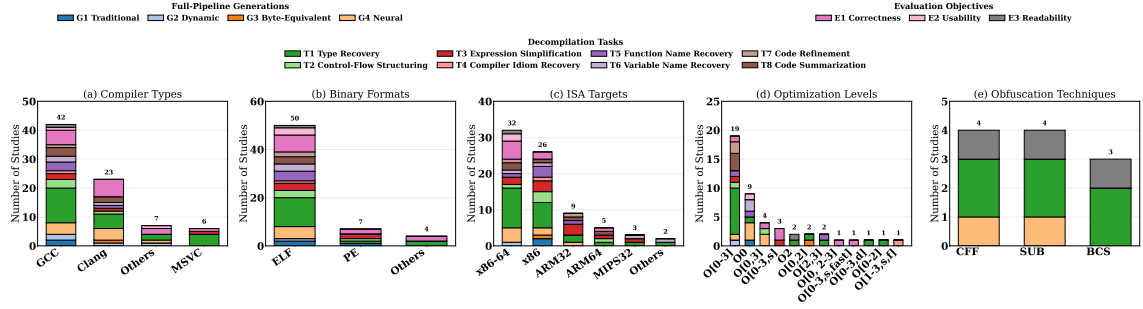}
        \caption{
            Distribution of compilation settings, 
            including compilers, 
            binary formats,
            ISA targets, 
            optimization levels, 
            and
            obfuscation techniques used in 
            experimental configurations.
       }
    \label{fig:evaluation_settings}
\end{figure*}

\subsection{Compiler Configuration}
\OA{
Compiler configuration plays a 
central role in shaping the
resulting binary, influencing its structure, optimization patterns,
debug information, and 
ultimately the difficulty of decompilation.
We therefore examine the distribution
of compilation settings
used across decompilation studies,
as shown in~\autoref{fig:evaluation_settings}.
}

\subsubsection{\OA{Compiler Types}}

Compilers produce different code-generation patterns that can
substantially affect binary structure and decompilation.
\OA{
Their diversity is therefore important for assessing whether an
approach generalizes across toolchains.
}
\autoref{fig:evaluation_settings}(a)
\OA{
shows that GCC~\cite{gcc} dominates, appearing in 40 studies,
followed by Clang~\cite{clang} in 20, while MSVC~\cite{msvc}
and other compilers each appear in six.
GCC spans G1, G2, and G4, all eight task-oriented techniques,
and E1--E3, with its use concentrated in T1, with 12 studies.
Clang is used most frequently in E1 and T1, with six and five
studies, respectively, while four of the six MSVC studies
target T1.
}
\OA{
The dominance of GCC and Clang provides broad coverage of
open-source toolchains, but the limited use of alternative
compilers leaves cross-toolchain generalization comparatively
underexplored.
}

\subsubsection{\OA{Binary Formats}}
\OA{Binary formats determine how executable code, symbols,
relocations, sections, and metadata are represented and accessed
during decompilation.
Format diversity therefore affects both preprocessing and the
evidence available to subsequent analyses.
}
\autoref{fig:evaluation_settings}(b)
\OA{
shows a strong concentration on ELF, which appears in 50 studies,
compared with seven using PE and four using other binary formats (\eg Mach-O).
ELF spans nearly all decompilation paradigms, tasks, and
evaluation objectives, with its use concentrated in T1,
with 11 studies, followed by E1, with seven.
PE appears across a smaller set of settings, including G1,
T1--T3, and E1.
}
\OA{
The dominance of ELF supports reproducible Linux-oriented
evaluation, but the substantially lower coverage of PE and other
formats limits evidence of cross-format generalization.
}

\subsubsection{Instruction Set Architecture Targets}

ISAs influence instruction encoding, calling conventions, and
architecture-specific patterns encountered during decompilation.
\OA{
ISA diversity is therefore important for determining whether a
technique generalizes beyond the architecture on which it was
developed or evaluated.
}
\autoref{fig:evaluation_settings}(c)
\OA{
shows that x86 and x86-64 dominate the surveyed evaluations,
while ARM32, ARM64, MIPS32, and other ISAs appear substantially
less frequently.
The x86 family is represented across full-pipeline approaches,
task-oriented techniques, and evaluation objectives, with T1
accounting for a substantial share of multi-ISA evaluation.
}
\OA{
This concentration provides extensive evidence for x86-family
binaries but comparatively weaker evidence of generalization
to ARM, MIPS, and other architectures.
}

\subsubsection{\OA{Optimization Levels}}

Compiler optimization transforms program structure and can
substantially alter the information available to a decompiler.
\OA{
Evaluating different optimization levels is therefore important
for determining robustness to compiler transformations.
}
\autoref{fig:evaluation_settings}(d)
\OA{
shows that optimization settings are evaluated across
full-pipeline approaches, task-oriented techniques, and
correctness evaluation, with T1 accounting for the largest
share.
The standard \texttt{-O0}--\texttt{-O3} range dominates,
with \texttt{-O0} and \texttt{-O3} being particularly common.
}
\OA{
These levels expose decompilers to substantially different
amounts of compiler transformation, but inconsistent reporting
and the additional use of settings such as \texttt{-Os},
\texttt{-Og}, and \texttt{-Of} complicate direct comparison
across studies.
}

\subsubsection{Obfuscation Techniques}

Obfuscation deliberately transforms binary structure to hinder
program analysis and therefore provides a challenging setting
for evaluating decompilation robustness.
\autoref{fig:evaluation_settings}(e)
\OA{
shows that obfuscation is evaluated in only four surveyed
studies.
Two target T1, while G4 and E3 account for one study each,
indicating substantially narrower coverage than other
compilation settings.
}
\OA{
The evaluated transformations primarily include compiler-level
techniques such as bogus control flow, control-flow flattening,
and instruction substitution.
}
\OA{
Consequently, existing evidence primarily reflects robustness
to a small set of compiler-level transformations rather than
stronger real-world obfuscation such as virtualization and
polymorphic transformations.
}

\section{Open Challenges and Future Research Directions}
\label{sec_challlenges}

\begin{figure*}[htbp]
    \centering
        \includegraphics[width=0.95\linewidth]{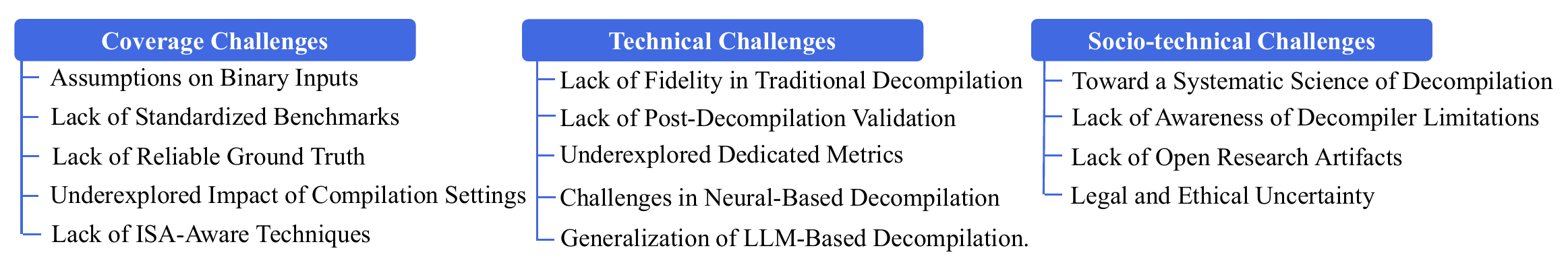}
        \caption{
            Key challenges in decompilation 
            research, organized into
            coverage, technical, and
            socio-technical dimensions. 
       }
    \label{fig:challenges}
\end{figure*}

In this section, we identified \OA{14}
critical open issues in binary decompilation, 
synthesized from our systematization and 
organized into coverage, 
technical, and 
socio-technical dimensions, 
as summarized in~\autoref{fig:challenges}.

\subsection{Coverage Challenges}
\OA{
Coverage refers to the
extent to which
evaluation settings represent
the diversity
of binaries encountered 
in practice.
Insufficient coverage limits
the ability to generalize 
reported results.
}

\subsubsection{\OA{Assumptions on Binary Inputs}}

\OA{
Most surveyed studies evaluate well-formed binaries produced by
standard compiler toolchains and generally assume that inputs are
unpacked and unobfuscated.
Consequently, packed executables, malformed binaries,
partially recovered firmware, JIT-generated code, and binaries
modified after compilation largely fall outside existing
evaluation settings.
Although some studies evaluate obfuscation using
OLLVM~\cite{pei2021stateformer,zhu2024tygr,eom2024r2i},
they assess robustness to transformed binaries rather than
deobfuscation itself.
Future studies should therefore report their input assumptions
explicitly to clarify the scope and applicability of their results.
}

\subsubsection{Lack of Standardized Benchmarks}

The absence of shared, standardized \OA{and representative} benchmarks
hinders reproducibility and fair comparison across decompilation tools.
\OA{
This problem concerns not only benchmark availability, but also
their representativeness and coverage of compilation effects.
For example, although GNU Coreutils~\cite{gnu2025coreutils}
is widely used, its programs share similar coding styles,
libraries, and implementation patterns and may therefore
represent only a narrow portion of real-world software.
Recent efforts such as Decompile-Bench~\cite{tan2025decompile}
improve scale and standardization through large collections of
binary--source pairs, but do not by themselves ensure controlled
coverage of individual compiler transformations.
For example, Abusabha \etal~\cite{abusabha2025deep}
show that inlining behavior can be systematically tuned to expose
substantially different binary structures.
Future benchmarks should therefore combine broad program coverage
with consistent granularity, explicit compilation metadata, and
controlled variation of compiler transformations.
}

\subsubsection{Lack of Reliable Ground Truth}

Defining reliable ground truth for decompilation remains an unsolved problem. Existing strategies—such as relying on DWARF debug information~\cite{pal2024len}, marker-based instrumentation~\cite{ahmed2021learning}, harnessed compilation pipelines~\cite{basque2024ahoy}, and modified compiler internals~\cite{katz2018using}—introduce substantial limitations in accuracy and realism. Debug information is often incomplete or inaccurate, especially in optimized binaries~\cite{li2020debug, di2021s}, while synthetic or artificially constrained compilation setups fail to reflect real-world variability. Establishing standardized, high-fidelity, and diverse ground truth datasets is vital for both evaluation and training.

\subsubsection{\OA{Underexplored Impact of Compilation Settings}}

\OA{
Current decompiler evaluations rarely isolate how individual
compilation settings affect the heuristics on which
decompilers rely.
Bin2Wrong~\cite{yang2025bin2wrong} begins to address this gap
by systematically varying compiler configurations, including
optimization levels and individual compilation flags, to study
their effects on decompiler correctness.
Its findings show that evaluating only standard optimization
levels can overlook failures triggered by specific compiler
transformations or settings.
Related compilation-space exploration techniques such as
BinTuner~\cite{ren2021unleashing}, although not evaluated on
decompilers, could be adapted to generate binaries that trigger
decompiler heuristic failures.
Future work could further use LLMs to generate targeted
compilation configurations and test cases that expose when
specific recovery heuristics succeed or fail.
}

\subsubsection{Lack of ISA-Aware Techniques}

ISA differences have a pronounced impact on both traditional and nerual-based decompilers. For example, \OA{Fu~\etal~\cite{fu2019coda} observe that} x86-64 binaries present challenges due to their verbose instruction set and implicit control-flow semantics, whereas architectures like MIPS expose branch conditions explicitly. Despite this, most decompilation approaches treat all ISAs uniformly or optimize for a single architecture. There is a pressing need for ISA-specific adaptation strategies and cross-ISA evaluation methodologies to improve the robustness and generalizability of decompilation techniques~\cite{kvroustek2013reconstruction}.

\subsection{Technical Challenges}
\OA{
Technical challenges concern the intrinsic
limitations of current decompilation
techniques, including their ability to
recover accurate, readable, and semantically
faithful source code.
Addressing these challenges is essential
for improving decompiler capability
and reliability.
}

\subsubsection{Lack of Fidelity in Traditional Decompilation}

Traditional decompilers primarily rely on 
static analysis pipelines, 
which often fall short in recovering high-level structure 
from optimized or obfuscated binaries. 
Type inference typically focuses on heap and stack variables, 
often neglecting global variables and custom data layouts~\cite{dramko2024taxonomy,zhu2024tygr,zou2024d}. 
Meanwhile,
control-flow structuring algorithms~\cite{yakdan2015no, basque2024ahoy} 
may prioritize readability over 
correctness or inflate code size~\cite{gussoni2020comb}. 
Consequently, outputs may appear plausible yet
embed subtle semantic inaccuracies, undermining trust in
decompilation results. 
Addressing this fidelity gap will require integrating 
static, dynamic, and ML–based techniques 
into modular architectures 
capable of validating and
refining their own outputs.

\subsubsection{\OA{Lack of Post-Decompilation Validation}}

\OA{
Current decompiler outputs may require additional validation
to assess whether they preserve the semantics of the original binary.
This issue becomes particularly important for post-decompilation
techniques such as code summarization and code refinement,
which further interpret or transform decompiler output.
However, semantic validation after decompilation remains
underexplored~\cite{mantovani2022convergence,
botacin2019revenge, han2023queryx, gorzynski2025fuzzflesh}.
Future work should investigate semantic-equivalence checking,
compiler-assisted validation, and feedback mechanisms that can
detect and correct discrepancies introduced during decompilation
or subsequent refinement.
}

\subsubsection{Underexplored Dedicated Metrics}
In our survey, we found only one dedicated metric
designed specifically for 
decompilation, R2I~\cite{eom2024r2i}, which targets readability. 
Beyond this, evaluations typically rely on
generic measures, 
which are not tailored to the unique
challenges of decompilation. 
While narrow, task-specific metrics can be 
useful in certain contexts, 
they may not generalize.
For example, R2I is unsuitable for 
malware analysis scenarios where 
semantic understanding outweighs code readability.
Developing additional dedicated metrics
would enable more precise benchmarking
and foster systematic tool improvement.

\subsubsection{\OA{Challenges in Neural-Based Decompilation}}

\OA{
Our analysis identifies several recurring limitations in
neural-based decompilation.
As discussed previously, generative
approaches enlarge the output space in which syntactic or
semantic errors may occur.
Moreover, we show that many
surveyed models operate under limited input contexts,
constraining their ability to process long functions and
low-level representations.
These findings motivate further work on semantic robustness
and scalable modeling of larger program contexts.
}

\subsubsection{\OA{Generalization of LLM-Based Decompilation}}

\OA{
Our survey shows that LLMs are increasingly adopted across
multiple decompilation tasks, including type recovery,
function naming, code summarization, and code refinement.
However, studies differ substantially in their underlying
models, prompting or fine-tuning strategies, input
representations, and evaluation settings.
This heterogeneity makes it difficult to determine whether
reported improvements generalize across models, tasks, and
experimental configurations.
Future work should therefore establish more systematic
cross-model and cross-task evaluation of LLM-based
decompilation.
}

\subsection{Socio-technical Challenges}
\OA{
Socio-technical challenges concern how decompilation techniques
interact with users and the research ecosystem, affecting
reproducibility, adoption, collaboration, and responsible use.
}

\subsubsection{\OA{Toward a Systematic Science of Decompilation}}
\OA{
Beyond the primary studies reviewed in this survey,
several community tools illustrate emerging support for
cross-decompiler comparison and artifact synchronization.
For example, mdec~\cite{mdec} and
Decompiler Explorer~\cite{dogbolt} support comparison across
decompilers, while BinSync~\cite{binsync} synchronizes
reverse-engineering artifacts across analysis platforms.
However, these capabilities provide limited insight into why a
decompiler succeeds or fails, or which recovery strategy is
reliable for a given binary.
Future work should therefore develop adaptive and measurable
decompilation workflows that expose uncertainty (\eg warnings) 
and failure
signals and select recovery strategies according to input
characteristics.
}

\subsubsection{Lack of Awareness of Decompiler Limitations}

\OA{
Decompiled outputs are used in security workflows, yet
downstream analyses may not explicitly account for
decompilation artifacts such as semantic inaccuracies,
type errors~\cite{soni2025benchmarking}, and control-flow mistakes~\cite{behner2025sok}.
Future reverse-engineering pipelines should therefore
incorporate error awareness, validation, and correction
mechanisms when consuming decompiled code.
}

\subsubsection{Lack of Open Research Artifacts}
Roughly 48\% of the studies we reviewed did not release their code, models, or datasets—highlighting a significant barrier to reproducibility in decompilation research. This lack of transparency hinders progress, limits fair comparison, and obstructs independent validation. Calling out and encouraging open practices should become a shared responsibility within the research community. Open-sourcing code, pretrained models, and standardized benchmarks is essential for fostering collaboration and accelerating innovation.

\subsubsection{Legal and Ethical Uncertainty}

Despite the long-standing debate around the legality of reverse engineering, we found no decompilation studies that systematically address the legal or ethical implications of their work. The legal status of decompilation varies significantly across jurisdictions, touching on issues such as copyright, fair use, and anti-circumvention laws~\cite{cambridge_ethics_decompilation}. Yet, this dimension is almost entirely absent from academic literature. Integrating those perspectives is essential to avoid unintentional misuse and to promote responsible, transparent research practices.

\section{Threats to Validity}\label{sec_threat}

As with any systematic review, this study 
is subject to several limitations, 
including potential biases in 
the search, selection, and classification processes. 
We mitigated these risks through 
iterative protocol refinement, 
pilot reviews, and a 
transparent methodology, 
following established SLR guidelines 
such as those proposed by
Kitchenham \etal~\cite{keele2007guidelines}.

\begin{itemize}

\item \textit{Search Bias.}
    The literature search spanned 
    from the earliest available 
    records through December \OA{2025}, 
    using publications 
    indexed in Google Scholar, 
    IEEE Xplore, and the ACM 
    Digital Library. 
    \OA{
    When publication years 
    differed across sources,
    we used the year reported
    by Google Scholar
    to maintain consistent
    chronological ordering.
    }
    Some relevant studies may have
    been overlooked due to limitations in
    search string expressiveness, 
    keyword coverage, or 
    indexing inconsistencies.
\item \textit{Selection Bias.}
    Borderline inclusion decisions 
    required subjective judgment, 
    which may have influenced the 
    final dataset and affected the
    representativeness of our sample.
\item \textit{Classification Bias,}
    Study categorization was performed 
    manually based on perceived contributions. 
    This introduces inherent 
    subjectivity 
    and may have led to 
    misclassification—particularly for 
    studies spanning multiple research domains.

\item \textit{Generalizability.}
    Our findings primarily target 
    C-language decompilation and 
    peer-reviewed academic literature. 
    As a result, conclusions 
    may not generalize
    to other ecosystems 
    (\eg Java or Python decompilation) or 
    to industrial settings not 
    represented in academic discourse.

\end{itemize}

\section{Conclusion}\label{sec_conclusion}

Decompilation continues to be a critical 
component of modern binary analysis, 
yet it remains an active and evolving research area.
Through this systematic review,
we synthesized developments across 
decompilation approaches, 
evaluation~\OA{capabilities}, 
and 
technical foundations—including benchmarks
and compilation settings—and 
identified 
\OA{14}
open challenges that 
frame the trajectory of future work.
Our analysis highlights the increasing
role of ML, and in particular LLMs, 
as promising directions for 
enabling more adaptive and
semantically informed decompilation.
We hope this survey serves as
a foundation for future research, 
fostering the development of 
more robust, transparent, and 
practically deployable decompiler systems.

\section{\OA{Ethical Concerns}}

\OA{
This survey combined manual analysis with AI-assisted tools.
NotebookLM was used to support literature organization and
data extraction, while ChatGPT assisted with language refinement,
manuscript organization, and consistency checking.
All study selection, classification, technical analysis, and
final decisions were conducted and verified by the first author.
}

\bibliographystyle{ACM-Reference-Format}
\bibliography{references}

\end{document}